\documentclass[pre,aps,twocolumn,showpacs,amsmath,amssymb,floatfix,superscriptaddress,citeautoscript]{revtex4-1}
\usepackage{graphicx}
\usepackage{dcolumn}
\usepackage{bm}
\usepackage{amsmath}
\usepackage{amssymb}
\usepackage{amssymb, verbatim}
\usepackage[normal]{subfigure}

\usepackage{hyperref}
\hypersetup{
        colorlinks=true,
}

\makeatletter

\usepackage{color}

\begin{document}

\title{Majorana zero modes choose Euler numbers - revealed by full counting statistics}

\author{Dong E. Liu}
\affiliation{Station Q, Microsoft Research, Santa Barbara, California 93106-6105, USA}
\author{Alex Levchenko}
\affiliation{Department of Physics and Astronomy, Michigan State
University, East Lansing, Michigan 48824, USA}
\author{Roman M. Lutchyn}
\affiliation{Station Q, Microsoft Research, Santa Barbara, California 93106-6105, USA}

\date{\today}

\begin{abstract}
We study transport properties of a quantum dot coupled to a Majorana zero mode and two normal leads. We investigate the full counting statistics of charge tunneling events which allows one to extract complete information about current fluctuations. Using a Keldysh path-integral approach, we compute the cumulant generating function. We first consider a noninteracting spinless regime, and find that for the symmetric dot-lead couplings, the zero-frequency cumulants exhibit a universal pattern of Euler numbers, independent of the microscopic parameters. For a spinful case, the Coulomb interaction effects are discussed for both strong interaction (single-electron occupancy regime) and weak interactions (perturbative regime). Compared to the case without Majorana coupling, we show that, while the tunneling conductance might exhibit zero-bias anomaly, the full counting statistics is qualitatively different in the presence of the Majorana coupling.
\end{abstract}


\maketitle

\section{Introduction}

Majorana zero-energy modes (MZMs) have recently attracted enormous theoretical and experimental
attention \cite{Reich,Brouwer_Science,Wilczek2012,AliceaRev} due to their exotic non-Abelian braiding statistics \cite{Moore1991,Nayak1996,ReadGreen} and potential application to fault-tolerant topological quantum computation \cite{TQCreview}. A large number of theoretical proposals has been put forward to realize MZMs in topological superconductors (TSCs)~\cite{Fu&Kane08,Fu&Kane09,Sau10,Alicea10,LutchynPRL10,1DwiresOreg,MajoranaTInanowires,SauNature'12,Nadj-Perge13}, see also reviews~\cite{AliceaRev, BeenakkerReview, NayakReview2015} for more details. One of the most promising proposals involves a semiconductor nanowire with strong spin-orbit interaction coupled to a conventional $s$-wave superconductor~\cite{LutchynPRL10,1DwiresOreg}. An appropriate combination of the spin-orbit coupling, Zeeman splitting and induced $s$-wave pairing allows one to realize an effectively spinless $p$-wave superconductivity at the interface which is characterized by the presence of MZMs bound to certain defects (i.e vortices in 2D and domain walls in 1D)~\cite{ReadGreen, 1DwiresKitaev}. The simplest way to detect MZMs is to measure local density of states at the defect and to probe the emergence of the zero-energy resonance across the topological phase transition. The first Majorana tunneling spectroscopy experiment, based on a semiconductor/superconductor heterostructure proposal \cite{LutchynPRL10,1DwiresOreg}, was performed in Delft \cite{Mourik2012}. Later on, the observation of zero bias peak in a finite magnetic field, consistent with the theoretical predictions~\cite{ZeroBiasAnomaly61}, was reported by many other experimental groups~\cite{Mourik2012, Das2012, Deng2012, Fink2012, Churchill2013,Nadj-PergeScience,Deng_arxiv2014}. The main challenge of these measurements is to exclude the other false-positive contributions to the zero-bias peak that are ubiquitous in condensed matter systems such as Kondo effect~\cite{Kondo_Aguado, Pillet'13}, disorder in the topological region\cite{liujie12, Neven13, LobosPRL13, Sau&DasSarma13, Hui14} and in the leads~\cite{Bagrets12, Pikulin12} as well as some other resonant Andreev scattering phenomena~\cite{NayakReview2015}. The feature distinguishing the Majorana origin of the zero-bias peak from the other mechanisms is the quantized zero-bias peak conductance of $2e^2/h$ which is a universal property of Majorana zero modes~\cite{Fidkowski2012, Lutchyn&Skrabacz13}. However, due to the large subgap conductance (so-called ``soft gap" problem) observed in tunneling experiments~\cite{Mourik2012, Das2012, Deng2012, Fink2012, Churchill2013,Nadj-PergeScience,Deng_arxiv2014}, the largest observed height of the zero-bias peak was at most 30$\%$ of the predicted value. Therefore, additional experimental tests
\cite{Rokhinson2012, Leijnse11,Liu11,Golub'11,LeePRB13,Vernek14,ChengPRX2014,
DasSarmainterferometry,Grosfeldinterferometry,Fu&Kandinterferometry,NilssonNoise,AkhmerovPRL09,Pikulin12,Rainis13,DissipativeMF,Valentini14, Cheng2015} are necessary in order to conclusively confirm the presence of MZMs in the semiconductor-superconductor heterostructures.

\begin{figure}
\centering
\includegraphics[width=3.4in,clip]{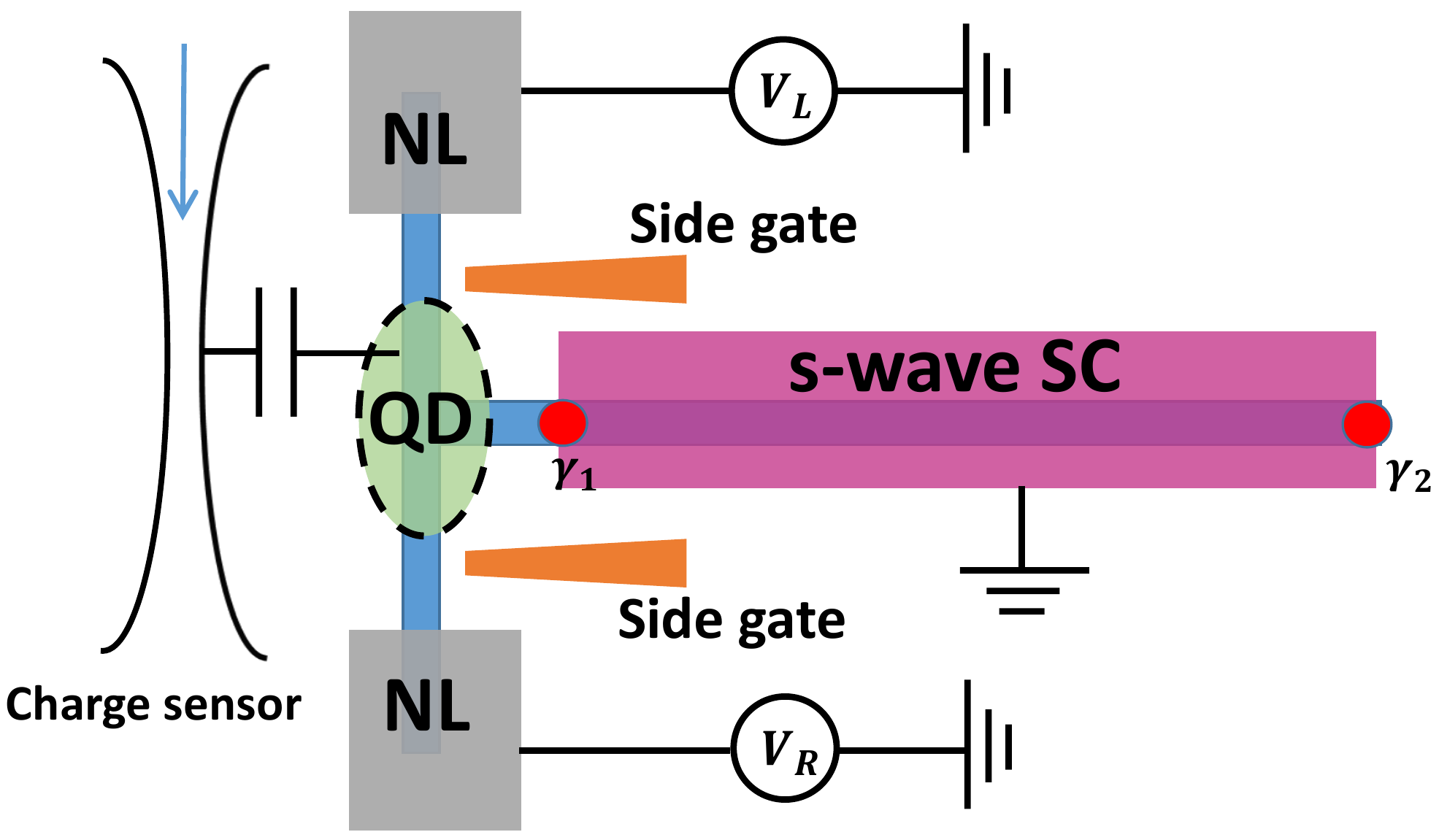}
\vspace{-0.1in}
\caption{Proposed experimental setup to measure the distribution function of the transmitted charge, namely the full counting statistics. The QD is created near the center of the semiconductor nanowire T-junction, two normal metal leads are attached to the upper and lower legs, and an $s$-wave superconductor is in proximity to the third lead. The latter realizes topological superconductor hosting two MZMs
$\gamma_1$ and $\gamma_2$. Another electron channel capacitively couples the T$-$junction, and can be used as a charge sensor to measure the charge distribution function.
}
\vspace{-0.2in}
\label{fig:setup}
\end{figure}

In this paper, we study current correlations in a mesoscopic device consisting of a quantum dot (QD) coupled to a MZM and two normal leads. The possibility to tune the couplings between QD and other conductors as well as QD gate voltage allows one to study current correlations in a well-controlled environment. We show here that in the case of a symmetric left-right normal metal coupling, see Fig.\ref{fig:setup} for a layout of the proposed device, current fluctuations are characterized by a universal pattern of the zero-frequency cumulants. We argue that the measurement of such cumulants allows one to exclude other false-positive signatures in tunneling transport and uniquely identify the presence of the putative Majorana modes.

The transport properties of a strongly interacting QD coupled to a MZM and a single normal lead (NL)  have been studied using master equations, valid in the high-temperature regime, in Ref.~\onlinecite{Leijnse11}. The low-temperature behavior of the MZM-QD-NL system and the interplay between Kondo and Majorana couplings was considered in Ref.~\onlinecite{Golub'11} finding that zero bias tunneling conductance exhibits strong temperature dependence, which is distinct from that of a MZM-NL structure~\cite{Fidkowski2012, Lutchyn&Skrabacz13}. Later on, Cheng, et al. \cite{ChengPRX2014} revisited the the low-temperature behavior of the MZM-QD-NL system, and found that Majorana coupling significantly modifies the low-energy properties of the QD and drives the system to a new (different from Kondo) infrared fixed point. They also confirmed that the temperature dependence of the zero bias conductance at the particle-hole symmetric point is similar to that of the MZM-NL structure \cite{Lutchyn&Skrabacz13}. The zero-bias conductance of a noninteracting QD with a side-coupled MZM (ungrounded TSC) through two normal leads was considered in Refs. \onlinecite{Liu11, Vernek14}, where it was predicted that the tunneling conductance is given by $e^2/2h$ for symmetric QD-lead couplings. Ref. \onlinecite{LeePRB13} considered a spinful QD in the Kondo regime for this two-lead structure, and studied the QD spectrum and zero-bias conductance by using numerical renormalization group method, which shows that the zero-bias conductance is $3e^2/2h$ for small QD-MZM coupling. The shot noise of a different two-lead structure (with a grounded TSC) has been studied in Ref. \onlinecite{Liu14} for both noninteracting spinless QD and spinful Kondo QD predicting that the shot noise not only shows universal behaviors but also can be used for qualitatively distinguishing MZMs with other modes. There has been also an experimental interest in QD-superconductor devices. The interplay of the Kondo effect and superconductivity has been revisited in Refs.~\cite{Kondo_Aguado, Lee_arxiv2013, Chang2013, Deng_arxiv2014}. A natural realization for the proposed experimental setup, see Fig. \ref{fig:setup}, involves a T-junction of the semiconductor nanowires which can be grown using vapour - liquid - solid growth technique~\cite{plissard2013}. The QD can be created near the junction, two normal leads and the TSC are connected to each leg of the T-junction. Thus, we believe that the setup we propose in the paper is within the experimental reach.

Although conductance and shot noise exhibit peculiar universal dependence due to the MZM coupling, it is insightful to obtain the full probability distribution function of the charge transferred through the QD, which can serve as the Majorana sensor. The theory of full counting statistics (FCS) \cite{LevitovJETP93,LevitovJMP96,NazarovFCSreview} for charge transport in mesoscopic systems was established by analyzing nonequilibrium transport.
A great effort has been made to investigate various aspects of
FCS in a variety of systems theoretically \cite{BelzigFCSSC01,BelzigFCSDiffusive,KindermannPRL03,BagretsPRL04,KomnikPRL05,Gogolin&Komnik06,Gogolin&KomnikPRL06,
UtsumiPRL06,SchmidtPRB07,EntanglementFCS,UrbanPRB10,SakanoPRB11,WeithoferFCSMZM,Bagrets&Levchenko} and experimentally
\cite{ReuletPRL03,BomzePRL05,GustavssonPRL06,FujisawaScience06,GershonPRL08,FlindtPNASFCS,UbbelohdeExperiment,MaisiPRL14}. Recently, the FCS calculation has also been considered for electron transport through multiterminal networks of MZMs \cite{WeithoferFCSMZM}. In this paper, we study FCS of charge tunneling through a QD with a side-coupled TSC, or equivalently a QD coupled to a MZM. The charge transport is measured between two normal leads. Here we assume that TSC is grounded so there is also Andreev current between (left) lead and the superconductor. Using the Keldysh path-integral approach~\cite{Kamenev}, we compute the cumulant generating function. We first consider a noninteracting spinless QD, and find that for the symmetric dot-lead couplings, the zero-frequency cumulants exhibit a universal pattern described by Euler polynomial. This result is independent of the microscopic parameters such as QD energy level and QD-MZM coupling. For a spinful QD with a small QD Zeeman splitting,
we compute FCS in the regime of weak (perturbative regime) and strong (single-electron occupancy regime) Coulomb interactions. In the former case, we compute the interaction-induced correction to the cumulant generating function up to the leading order in $U$ (i.e. $U^2$). In the latter case, we apply a slave boson mean field approach, and study the FCS due to the interplay between the Kondo and Majorana coupling.

The paper is organized as follows. In Sec. \ref{sec:formalism}, we review the formalism of the FCS calculation for the mesoscopic transport problem. In Sec. \ref{sec:NonInteracting}, we introduce the QD-MZM model and compute the cumulant generating function for this model with noninteracting spinless QD. In Sec. \ref{sec:Perturbation}, we consider weak Coulomb interaction effect for a spinful QD, and compute the leading order interaction correction to the cumulant generating function by using a diagrammatic perturbation method. In Sec. \ref{sec:Kondo}, we consider strong Coulomb interaction effect for a spinful QD, and study how Kondo and Majorana couplings affect the FCS within a slave boson mean field approach. Finally, the conclusions are shown in the Sec. \ref{sec:conclusion}.

\section{Full Counting Statistics: General formalism}\label{sec:formalism}

In this section, we will review the formalism for calculating full counting statistics (FCS) of charge fluctuations in a mesoscopic system, we refer a reader to Ref.~\cite{BagretsFCSReview} for more details. Consider the distribution function $\mathcal{P}_{q}$ for $q$ electrons to be transferred through a mesoscopic device within the measurement time $\mathcal{T}$. Here we assume that the measurement time is long enough ($\mathcal{T}\gg e/I$) so that the average number of electrons $M$ transferred within $\mathcal{T}$ is large, i.e. $M \gg 1$. The distribution function $\mathcal{P}_{q}$ allows one to extract more information about the nature of the charge carriers as well as the statistics of the charge fluctuations. In particular, tails of the distribution contain information about the statistics of rare events. From the theoretical point of view, rather than $\mathcal{P}_{q}$, it is more convenient to compute the cumulant generating function (CGF) $\chi(\lambda)$, defined by a Fourier transform
\begin{equation}
 \chi(\lambda)=\sum_{q} e^{i q \lambda} \mathcal{P}_{q}.
\end{equation}
Here the auxiliary variable $\lambda$ represents a counting field. From the CGF, one can calculate the cumulants
$\langle\!\langle \delta^n q \rangle\!\rangle$ (irreducible moments of $\mathcal{P}_{q}$)
\begin{equation}
 \langle\!\langle \delta^n q \rangle\!\rangle = \frac{\partial^{n}}{\partial(i\lambda)^n} \rm{ln} \chi(\lambda) \Big|_{\lambda=0}.
\end{equation}
Thus, the average current and zero-frequency symmetrized current noise can be obtained by a simple differentiation:
\begin{align}
 \!I &= \frac{1}{\mathcal{T}}\int_{0}^{\mathcal{T}} dt \langle \hat{I}(t) \rangle =
               \frac{-i}{\mathcal{T}}\frac{\partial \rm{ln} \chi}{\partial \lambda}\Big|_{\lambda=0},\label{eq:current_def}\\
 \!S &=\int_{0}^{\mathcal{T}} dt \langle \delta\hat{I}(t) \delta\hat{I}(0)+\delta\hat{I}(0) \delta\hat{I}(t)\rangle =
                \frac{-1}{\mathcal{T}}\frac{\partial^2 \rm{ln} \chi}{\partial \lambda^{2}}\Big|_{\lambda=0}.\label{eq:shot_def}
\end{align}
The third cumulant and the fourth cumulant describe the asymmetry (or skewness) and the kurtosis (or sharpness) of the distribution function, and can be also straightforwardly obtained using $\chi(\lambda)$.

In order to define the CGF in a proper way, Levitov and Lesovik introduced a ``gedanken scheme'', in which a spin-$1/2$ system magnetically couples to the electric current \cite{LevitovJETP93,LevitovJMP96}. Based on their definition, the CGF in a Keldysh formalism is given by
\begin{equation}
 \chi (\lambda)=\Big\langle T_C \exp\big[-i\int_{C} dt H^{\lambda}(t) \big]    \Big\rangle
\end{equation}
with
\begin{equation}\label{eq:Hlambda}
 H^{\lambda}(t) = H + \frac{\lambda(t)}{2} \hat{I}.
\end{equation}
Here, the integration is preformed along the Keldysh contour $C$, $T_C$ is time ordering;
$\langle\cdots\rangle$ is the quantum-mechanical average~\cite{Kamenev}. The second term in Eq.\eqref{eq:Hlambda} describes the interaction between electron current and counting field, and $\lambda(t)$ has different sign on the two branches of the Keldysh contour, i.e. $\lambda(t)=\lambda^{(1)}$
for forward branch and $\lambda(t)=\lambda^{(2)}$ for backward branch with $\lambda^{(1)}=-\lambda^{(2)}=\lambda$.

\section{FCS for a spinless QD with side-coupled MZM}\label{sec:NonInteracting}

\subsection{Theoretical Model}

We now consider the setup shown in Fig. \ref{fig:setup} - a quantum dot (QD) is coupled to an end of a one-dimensional (1D) topological superconductor (TSC) hosting two Majorana zero modes (MZMs) $\gamma_1$ and $\gamma_2$ at the opposite ends. For pedagogical reasons, we first consider the spinless model for this setup (assuming that the Zeeman splitting is very large), and relegate the discussion of the spinful case to the next sections. In the former case, the model is essentially noninteracting, and one can calculate FCS exactly. The QD is coupled to two normal spinless leads which can be used for transport measurements. Given that each lead couple to the QD at a single point, one can perform unfolding transformation and reduce the problem to the one corresponding to a quantum impurity coupled to one dimensional free fermions. Thus, the corresponding Hamiltonian reads
\begin{equation}
 H=H_{\rm Lead}+H_{\rm QD-MZM} + H_{\rm T},
 \label{eq:H_spinless}
\end{equation}
where the Hamiltonians for the leads, QD-MZM system, the Lead-QD couplings are respectively given by
\begin{eqnarray}
 &H_{\rm Lead} = -i v_F \sum_{\alpha=L,R} \int dx \psi_{\alpha}^{\dagger}(x) \partial_x \psi_{\alpha}(x),\\
 &H_{\rm QD-MZM} = \epsilon_d d^{\dagger}d+ i\kappa (d+d^{\dagger})\gamma_1+ i\delta\gamma_1 \gamma_2 , \\
 &H_{\rm T} = \sum_{\alpha=L,R} \Big( t_{\alpha}\psi_{\alpha}^{\dagger}(0) d + t_{\alpha}^{*} d^{\dagger}\psi_{\alpha}(0)   \Big)
\end{eqnarray}
Here $\psi_{\alpha}^{\dagger}$ ($d^{\dagger}$) is creation operator for an electron in the $\alpha$-lead (QD), $\epsilon_d$ is the energy level in the QD, and $t_{\alpha k}$ ($\kappa$) is the tunnel coupling between the leads (MZM) and the QD. The effective Hamiltonian for the TSC is given in terms of the low-energy degrees of freedom (MZMs) assuming that the induced superconducting gap $\Delta$ is the largest energy scale in the problem. For a finite-length $L$ TSC, the coupling $\delta$ between two MFs is
exponentially small $\delta\sim \Delta \exp(-L/\xi)$ with the coherence length being $\xi=v_F/\Delta$.

We now derive the CGF for our QD-MZM model assuming the symmetric source-drain bias ($\mu_L=eV/2$ and $\mu_R=-eV/2$). The Hamiltonian including the counting field can be written as
\begin{equation}
 H^{\lambda}(t) = H + \sum_{\alpha=L,R}\frac{\lambda_{\alpha}(t)}{2} \hat{I}_{\alpha}.
\end{equation}
where the current operator for the $\alpha$-junction is $\hat{I}_{\alpha}=ie [H,N_{\alpha}]$ with the electron number operator $N_{\alpha}$ for the $\alpha$ lead. One can apply a gauge transformation to remove the last term in the $H^{\lambda}(t)$, and obtain
\begin{eqnarray}\label{eq:Hamiltonian_counting}
 H^{\lambda} &=& H_{\rm Lead}+H_{\rm QD-MZM} + H_{\rm T}^{\lambda}, \\
  H_{\rm T}^{\lambda} &=&  \sum_{\alpha=L,R} \Big( t_{\alpha} e^{-i\lambda_{\alpha}(t)/2}
  \psi_{\alpha}^{\dagger}(0) d + h.c. \Big).
\end{eqnarray}
We note that for $\kappa=0$, the gauge symmetry of the Hamiltonian allows one to gauge away one of the counting fields so it is enough to keep the counting field in one of the junction. In the general case (i.e $\kappa\neq 0$), however, we need to keep both counting fields $\lambda_{\alpha}$.

We can now compute the path integral for the effective action defined by the Hamiltonian~\eqref{eq:Hamiltonian_counting}. Given that the presence of superconductor (i.e. MZM coupling) breaks particle number conservation, the QD Green function contains anomalous contributions,
e.g. $\langle T_C d(t) d(t') \rangle\neq 0$. Therefore, we introduce
Nambu spinors: $\vec{\Psi}_{\alpha}^{\dagger}=(\psi_{\alpha}^{\dagger}, \psi_{\alpha})/\sqrt{2}$
and $\vec{\Psi}_{d}^{\dagger}=(d^{\dagger}, d)/\sqrt{2}$, where $\alpha=L, R$ is the lead index. The effective Keldysh action now reads
\begin{equation}
  S = S_{\rm Leads}+ S_{\rm QD-MZM}+ S_{\rm T},
\end{equation}
where
\begin{eqnarray}
&&S_{\rm Leads} = \sum_{\alpha} \int_{C}\int_{C} dtdt' \vec{\Psi}_{\alpha}^{\dagger}(t)
               \breve{Q}_{0,\alpha}^{-1}(t,t') \,\vec{\Psi}_{\alpha}(t')\nonumber\\
&&S_{\rm QD-MZM}     =  \int_{C}\int_{C} dtdt' \vec{\Psi}_{d}^{\dagger}(t)\, \breve{Q}_{0,dd}^{-1}(t,t')\, \vec{\Psi}_{d}(t'),\\
&&S_{\rm T} = -\sum_{\alpha}\int_{C} dt \Big( t_{\alpha} e^{-i\frac{\lambda_{\alpha}(t)}{2}} \psi_{\alpha}^{\dagger} d+c.c.  \Big)\\
&&\quad\quad  =  -\sum_{\alpha}\int_{C} dt (\vec{\Psi}_{\alpha}^{\dagger}(t) M_{T,\alpha} \vec{\Psi}_{d}(t)+h.c.),
\end{eqnarray}
are the actions for leads, QD, and Lead-QD coupling, and
\begin{equation}
 M_{T,\alpha}=
 \begin{pmatrix}
  t_{\alpha } e^{-i\frac{\lambda_{\alpha}(t)}{2}}&0
  \\ 0 & -t_{\alpha}^{*} e^{i\frac{\lambda_{\alpha}(t)}{2}}
 \end{pmatrix}.
\end{equation}
Here we have already integrated out the bulk degrees of freedom in the leads and kept only the field  $\vec{\Psi}_{\alpha}^{\dagger}(t)$ at the $x=0$, i.e. at the QD. The free lead Green's function $\breve{Q}_{0,\alpha}$ at $x=0$ in the Nambu space $\mathbb{N}$ can be written as
\begin{equation}
 \breve{Q}_{0,\alpha}(\omega) =
 \begin{pmatrix}
  g_{\alpha}^0(\omega) &0  \\
  0 & \widetilde{g}_{\alpha}^0(\omega)
 \end{pmatrix}
 \label{eq:Q0leads}
\end{equation}
where $\widetilde{g}_{\alpha}^0(t-t')$ is the P-H conjugation of $g_{\alpha}^0(t-t')$.
We perform Larkin-Ovchinnikov (L-O) rotation, and the Green function in Keldysh space becomes
\begin{eqnarray}
 g_{\alpha}^0(\omega) &=&-i\pi \rho_F
 \begin{pmatrix}
  1 & 2(1-2 n_{\alpha})\\0 & -1
 \end{pmatrix},\\
\widetilde{g}_{\alpha}^0(\omega) &=& -i\pi \rho_F
\begin{pmatrix}
  1 & 2(1-2 \widetilde{n}_{\alpha})\\0 & -1
 \end{pmatrix}.
\end{eqnarray}
One notices that $g_{\alpha}^{0,R}(\omega)=-\widetilde{g}_{\alpha}^{0,A}(-\omega)$
and $g_{\alpha}^{0,K}(\omega)=-\widetilde{g}_{\alpha}^{0,K}(-\omega)$.
Here, $n_{\alpha}$ is the Fermi distribution function of the $\alpha$ lead with
chemical potential $\mu_{\alpha}$, and $\widetilde{n}_{\alpha}$ corresponds
to the Fermi distribution function with $-\mu_{\alpha}$.
Assuming the symmetric source-drain bias $\mu_L=eV/2$ and $\mu_R=-eV/2$, one can relate the Fermi function for particles and holes
$\widetilde{n}_L=n_R$ and $\widetilde{n}_R=n_L$.
The free QD Green function (with MZM coupling) can be written as
\begin{equation}
 \breve{Q}_{0,dd}(\omega) =
 \begin{pmatrix}
  G_{0,d\bar{d}}^R & G_{0,d\bar{d}}^K & F_{0,dd}^R & F_{0,dd}^K \\
  0 & G_{0,d\bar{d}}^A & 0 & F_{0,dd}^A \\
   F_{0,\bar{d}\bar{d}}^R & F_{0,\bar{d}\bar{d}}^K & G_{0,\bar{d}d}^R & G_{0,\bar{d}d}^K \\
  0 & F_{0,\bar{d}\bar{d}}^A & 0 & G_{0,\bar{d}d}^A \\
 \end{pmatrix},
\label{eq:Q0dd}
\end{equation}
where the retarded components read \cite{Leijnse11,Liu14}
\begin{align}
 G_{0,d\bar{d}}^R(\omega) &= \frac{\omega+i\eta_S+\epsilon_d-\Sigma_{\rm M}(\omega)}{(\omega+i\eta_S-2\Sigma_{\rm M}(\omega))(\omega+i\eta_S)-\epsilon_d^2}\label{eq:G0dd1}\\
G_{0,\bar{d}d}^R(\omega) &= \frac{\omega+i\eta_S-\epsilon_d-\Sigma_{\rm M}(\omega)}{(\omega+i\eta_S-2\Sigma_{\rm M}(\omega))(\omega+i\eta_S)-\epsilon_d^2}\label{eq:G0dd2}\\
F_{0,dd}^R(\omega) &= F_{0,\bar{d}\bar{d}}^R(\omega)=\frac{-\Sigma_{\rm M}(\omega)}{(\omega+i\eta_S-2\Sigma_{\rm M}(\omega))(\omega+i\eta_S)-\epsilon_d^2}.\label{eq:F0dd}
 \end{align}
Here the the self-energy due to MZM coupling is $\Sigma_{\rm M}(\omega)=\kappa^2\omega/(\omega^2-\delta^2)$,
and the infinitesimal $\eta_S\rightarrow 0$. The Keldysh components are proportional to $\eta_S$, and, thus, can be set to zero.
After L-O rotation, the action for the tunneling part becomes
\begin{eqnarray}
 S_{\rm T} &=& -\sum_{\alpha} \int_{-\infty}^{\infty} dt \Big [
   \vec{\Psi}_{\alpha}^{\dagger} \big( \sum_{i=cl,q} M_{T,\alpha}^i\otimes \gamma^i \big) \vec{\Psi}_{d} \nonumber\\
    && + \vec{\Psi}_{d}^{\dagger} \big( \sum_{i=cl,q} M_{T,\alpha}^i\otimes \gamma^i \big)^{\dagger} \vec{\Psi}_{\alpha} \Big],
\end{eqnarray}
where
\begin{eqnarray}
  M_{T,\alpha}^{cl}&=&
  \begin{pmatrix}
   \frac{e^{-i\lambda_{\alpha}^{(1)}}+e^{-i\lambda_{\alpha}^{(2)}}}{2}&0 \\
   0& -\frac{e^{i\lambda_{\alpha}^{(1)}}+e^{i\lambda_{\alpha}^{(2)}}}{2}
  \end{pmatrix},\label{eq:TunningM11}\\
  M_{T,\alpha}^{q}&=&
  \begin{pmatrix}
   \frac{e^{-i\lambda_{\alpha}^{(1)}}-e^{-i\lambda_{\alpha}^{(2)}}}{2}&0 \\
   0& -\frac{e^{i\lambda_{\alpha}^{(1)}}-e^{i\lambda_{\alpha}^{(2)}}}{2}
  \end{pmatrix}.
  \label{eq:TunningM}
\end{eqnarray}
are written in the Nambu space whereas $\gamma^{cl}=\mathbb{I}$ and $\gamma^{q}=\sigma_1$ represent the Keldysh space. Note the relationship  $\lambda_{\alpha}^{(1)}=-\lambda_{\alpha}^{(2)}=\lambda_{\alpha}$ which allows one to simplify the expressions. After some manipulations,
the cumulant generating function can be written as
\begin{equation}
 \chi (\lambda)=\int D[d^{\dagger},d] D[\psi_{\alpha}^{\dagger},\psi_{\alpha}] e^{i (S_{\rm Leads}+S_{\rm QD-MZM}+S_{\rm T})}.
\end{equation}
Next, we perform Gaussian integration to find
\begin{equation}
 \ln \chi (\lambda) = \frac{\mathcal{T}}{2} \int_{-\infty}^{\infty} \frac{d \omega}{2\pi} \ln\left[
 \frac{\det\Big[ \mathbb{\breve{I}}_{4\times 4}-\breve{Q}_{0,dd} \breve{Q}_{M\lambda}}
               {\det\Big[ \mathbb{\breve{I}}_{4\times 4}-\breve{Q}_{0,dd} \breve{Q}_{M0}]}\right],
               \label{eq:generatingU0}
\end{equation}
where
\begin{equation}
\breve{Q}_{M\lambda}=\sum_{\alpha} \big( \sum_{i} M_{T,\alpha}^i\otimes \gamma^i \big)^{\dagger}
               \breve{Q}_{0,\alpha} \big( \sum_{i} M_{T,\alpha}^i\otimes \gamma^i \big) \Big].
\end{equation}
Eqs.\eqref{eq:generatingU0} is a general expression for the CGF. In the next sections, we will explicitly evaluate $\chi(\lambda)$ for different limiting cases.

\subsection{Results and Discussions}

\subsubsection{QD coupled to two normal leads.}

It is instructive to review first a simple case of a noninteracting QD coupled to two normal leads.
Taking the limit $\kappa=0$ in Eq. \eqref{eq:Q0dd} and substituting it into Eq. (\ref{eq:generatingU0}), one obtains
\begin{equation}\label{eq:CGM_Gogolin}
\ln \chi =\frac{\mathcal{T}}{2} \int_{-\infty}^{\infty} \frac{d \omega}{2\pi} \ln[(1+\Upsilon_+)(1+\Upsilon_-)]
\end{equation}
with the functions $\Upsilon_{\pm}$ being defined as
\begin{align}
\Upsilon_{\pm}&=\frac{4\Gamma_L\Gamma_R}{(\omega\pm\epsilon_d)^2+(\Gamma_L+\Gamma_R)^2}\\
&\times[n_L(1-n_R)(e^{i(\lambda_L-\lambda_R)}-1)+{R\leftrightarrow L}].\nonumber
\end{align}
The term in the second bracket of the logarithm function in Eq.\eqref{eq:CGM_Gogolin} is the particle-hole conjugation of the term in the first bracket. Since we consider a symmetric source-drain bias ($eV_L=-eV_R=eV/2$), the transformation $\omega\rightarrow -\omega$ (e.g. for the terms in the second bracket) will result in the following changes: $n_L\rightarrow 1-n_R$ and $n_R\rightarrow 1-n_L$. Thus, one can see that Eq.\eqref{eq:CGM_Gogolin} is consistent with the results of Ref.~\onlinecite{Gogolin&Komnik06}. Indeed, at zero temperature and to the linear order in applied bias $eV$, one obtains the well-known result for the shot noise in a QD:
\begin{eqnarray}
 \frac{S_{\rm LL}}{eV} &=& \frac{e^2}{h} (-i)^2 \frac{\partial^2 \ln \chi (\lambda_L,\lambda_R=0)}{\partial\lambda_L^{2}}\Big|_{\lambda_L=0}\nonumber\\
           &=& \frac{2e^2}{h} \frac{4\Gamma_L\Gamma_R [(\Gamma_L-\Gamma_R)^2+\epsilon_d^2]}{[\epsilon_d^2+(\Gamma_L+\Gamma_R)^2]^2}.
\end{eqnarray}
One can notice that the shot noise (as well as other cumulants) generically depend on the microscopic parameters for the QD such as, for example, $\epsilon_d$. Furthermore, in the resonant case corresponding to $\epsilon_d=0$ and $\Gamma_L=\Gamma_R$, the first eight cumulants are given by
\begin{eqnarray}
 &&\left\{C_1(0),\cdots C_8(0) \right\} =
 \{1, 0,0,0,0,0,0,0\},
   \label{eq:cumulants_NoMZM}
\end{eqnarray}
with $C_n(0)$ being defined as
\begin{equation}
 C_n(0)=\frac{\langle\langle \delta^n q \rangle\rangle}{M}.
\end{equation}
Notice that at the symmetric point $\Gamma_R=\Gamma_L$, all higher order ($n>1$) cumulants become zero as $\epsilon_d\rightarrow 0$. This dependence on $\epsilon_d$ is a generic feature because density of states in QD strongly depends on the gate voltage controlling $\epsilon_d$. As we show below, this is not the case when QD is coupled to a TSC.

\subsubsection{QD coupled to two normal leads and a TSC}\label{sec:spinlessQD_MZM}

Let us now consider a QD coupled to a TSC through MZM coupling, i.e. $\kappa\neq 0$. Substituting Eqs. (\ref{eq:Q0leads}),  (\ref{eq:Q0dd}),
(\ref{eq:TunningM11}), and (\ref{eq:TunningM}) into Eq. (\ref{eq:generatingU0}), we find
the following expression for the cumulant generating function:
\begin{widetext}
 \begin{eqnarray}
 \ln \chi (\lambda) &=& \frac{\mathcal{T}}{2} \int_{-\infty}^{\infty} \frac{d \omega}{2\pi} \ln
   \Big[ 1- \frac{\mathbb{C}_1}{\mathbb{K}(\lambda=0)}n_L (1-n_L) -\frac{\mathbb{C}_2}{\mathbb{K}(\lambda=0)}n_R (1-n_R)
   +\frac{\mathbb{B}_1}{\mathbb{K}(\lambda=0)}n_L (1-n_R)  \nonumber\\
   &&\quad\quad\quad  +\frac{\mathbb{B}_2}{\mathbb{K}(\lambda=0)}n_R (1-n_L)
    +\frac{\mathbb{F}}{\mathbb{K}(\lambda=0)}n_L n_R (1-n_L) (1-n_R)+\frac{\mathbb{J}}{\mathbb{K}(\lambda=0)}n_L n_R (n_L-n_R) \Big].
\label{eq:CGF_MZM_finiteT}
\end{eqnarray}
\end{widetext}
The coefficients ($\mathbb{C}_1$, $\mathbb{C}_2$, $\mathbb{B}_1$, $\mathbb{B}_2$, $\mathbb{F}$, $\mathbb{J}$, $\mathbb{K})$ in Eq.\eqref{eq:CGF_MZM_finiteT} are defined in the Appendix-\ref{app:CGF_coefficients}. Above expression can be simplified in the zero temperature limit where the terms proportional to $n_L (1-n_L)$, $n_R (1-n_R)$, $n_L n_R (1-n_L) (1-n_R)$, and $n_L n_R (n_L-n_R)$ vanish, and the corresponding expression for $\ln \chi (\lambda)$ becomes
\begin{align}
 \ln \chi (\lambda)\Big|_{T\rightarrow 0}&=\frac{\mathcal{T}}{2} \int_{-\infty}^{\infty} \frac{d \omega}{2\pi} \ln\Big[ 1 +\mathbb{N}_1 n_L (1-n_R)\nonumber\\
    &+\mathbb{N}_2 n_R (1-n_L) \Big]
\end{align}
where $\mathbb{N}_1(\omega)=\mathbb{B}_1/\mathbb{K}(\lambda=0)$ and $\mathbb{N}_2(\omega)=\mathbb{B}_2/\mathbb{K}(\lambda=0)$.
In addition, at $T=0$ we have $\big( n_L (1-n_R) \big)^i=n_L (1-n_R)$, $\big( n_R (1-n_L) \big)^i=n_R (1-n_L)$, and
$\big( n_L (1-n_R) \big)^i \big( n_R (1-n_L) \big)^j =0$ (if $i,j\neq 0$). Assuming $\mu_L>\mu_R$,
the generating function can be further simplified to
\newpage
\begin{eqnarray}
 \ln \chi (\lambda)\Big|_{T\rightarrow 0} &=& \frac{\mathcal{T}}{2} \int_{-\frac{eV}{2}}^{\frac{eV}{2}} \frac{d \omega}{2\pi}
     \ln\left(1+\frac{\mathcal{N}(\omega)}{\mathcal{D}(\omega)}\right).\label{eq:CGF_omega}
\end{eqnarray}
where the functions are $ \mathcal{N}(\omega)$ and $\mathcal{D}(\omega)$ are defined as
\begin{align}
\mathcal{N}(\omega)&=e^{-2 i \lambda _R} \Big\{8 e^{i \left(\lambda _L+\lambda _R\right)} \Gamma _L \Gamma _R
         \big[\epsilon_d^2+(\Sigma_{\rm M}(\omega) -\omega )^2\nonumber\\
         &+\left(\Gamma _L-\Gamma_R
            \right){}^2\big] + 4 \left(\Sigma_{\rm M}(\omega)^2+4 e^{2 i \lambda _L} \Gamma _L^2\right) \Gamma _R^2\\
          & -4 e^{2 i \lambda _R} \big[\left(1\!-\!e^{2 i \lambda
           _L}\right) \Sigma_{\rm M}(\omega)^2 \Gamma _L^2\!+\!2 \Gamma _L^3 \Gamma _R\nonumber\\
          &\!+\!\Sigma_{\rm M}(\omega)^2 \Gamma _R^2+2 \Gamma _L \Gamma _R \left(\epsilon_d^2+(\Sigma_{\rm M}(\omega) -\omega
                   )^2+\Gamma _R^2\right)\big]\Big\},\nonumber\\
\mathcal{D}(\omega) &= \big[\epsilon_d^2+(2 \Sigma_{\rm M}(\omega) -\omega ) \omega \big]^2+\left(\Gamma _L+\Gamma _R\right){}^2\\
&+\left(\Gamma _L+\Gamma _R\right)^2 \big[2 \left(\epsilon_d^2+2 \Sigma_{\rm M}(\omega)
^2-2 \Sigma_{\rm M}(\omega)  \omega +\omega ^2\right)\big].\nonumber
\end{align}
Eq.\eqref{eq:CGF_omega} is the main result of this section which allows one to compute cumulants as a function of various physical parameters. We now simplify above expression in the limit $\delta=0$ and small bias $eV\rightarrow 0$. We keep only the leading order terms in $eV$, and simply set $\omega=0$ in the integrand. \footnote{Note that limits $\delta \rightarrow 0$ and $\omega\rightarrow 0$ are non-commutative. Here we first set $\delta=0$ and then take the limit $\omega \rightarrow 0$}. After the simplifications, we arrive at a very simple expression for the CGF:
\begin{equation}\label{eq:spinless_cumulant}
   \frac{ \ln \chi}{M}\Bigg|_{eV\rightarrow 0} = \ln\left(\frac{\Gamma_L e^{i\lambda_L}+\Gamma_R e^{- i\lambda_R}}{\Gamma_L+\Gamma_R}\right),
\end{equation}
where $M=\mathcal{T}V/2\pi=\mathcal{T}Ve^2/h$ is the number of incoming particles during the waiting time. As mentioned above, the expression \eqref{eq:spinless_cumulant} only depends on the ratio of $\Gamma_L/\Gamma_R$, and is independent of many other microscopic parameters such as $\epsilon_d$ and $\kappa$. This universality is due to the finite density of states at zero energy induced by the Majorana leaking into the QD, and is a characteristic feature of topological superconductivity.

To get some insight we compute now currents through the left and right junctions. Using Eq.\eqref{eq:current_def}, one finds
\begin{align}
I_L&=\frac{e^2}{h}\frac{\Gamma_L}{\Gamma_R+\Gamma_L} V\\
I_R&=-\frac{e^2}{h}\frac{\Gamma_R}{\Gamma_R+\Gamma_L} V
\end{align}
Clearly, when $\Gamma_L \neq \Gamma_R$, there is Andreev contribution to the current due to the presence of a grounded superconductor:
\begin{align}
I_A=\frac{e^2}{h}\frac{\Gamma_R-\Gamma_L}{\Gamma_R+\Gamma_L}V.
\end{align}
One can notice that when $\Gamma_R=0$, we recover the previous results~\cite{ChengPRX2014} corresponding to a single lead coupled to a TSC. Indeed, given that the voltage drop between left lead and TSC is $V/2$, linear differential conductance $dI/dV$ is equal to the universal value of $2e^2/h$. Next, at the symmetric point $\Gamma_R=\Gamma_L$, Andreev current becomes zero, and linear differential conductance between right-left leads is $dI_L/dV={e^2}/{2h}$
which is consistent with the previous work on QD coupled to an ungrounded TSC~\cite{Liu11, Vernek14}. In addition, if we reverse the right lead voltage $V_R=-V/2 \longrightarrow V/2$, the linear conductance is also equal to the universal value $2e^2/h$ which is expected based on the RG analysis~\cite{Fidkowski2012, ChengPRX2014}.  

\begin{figure*}
 \centering
\includegraphics[width=5.0in,clip]{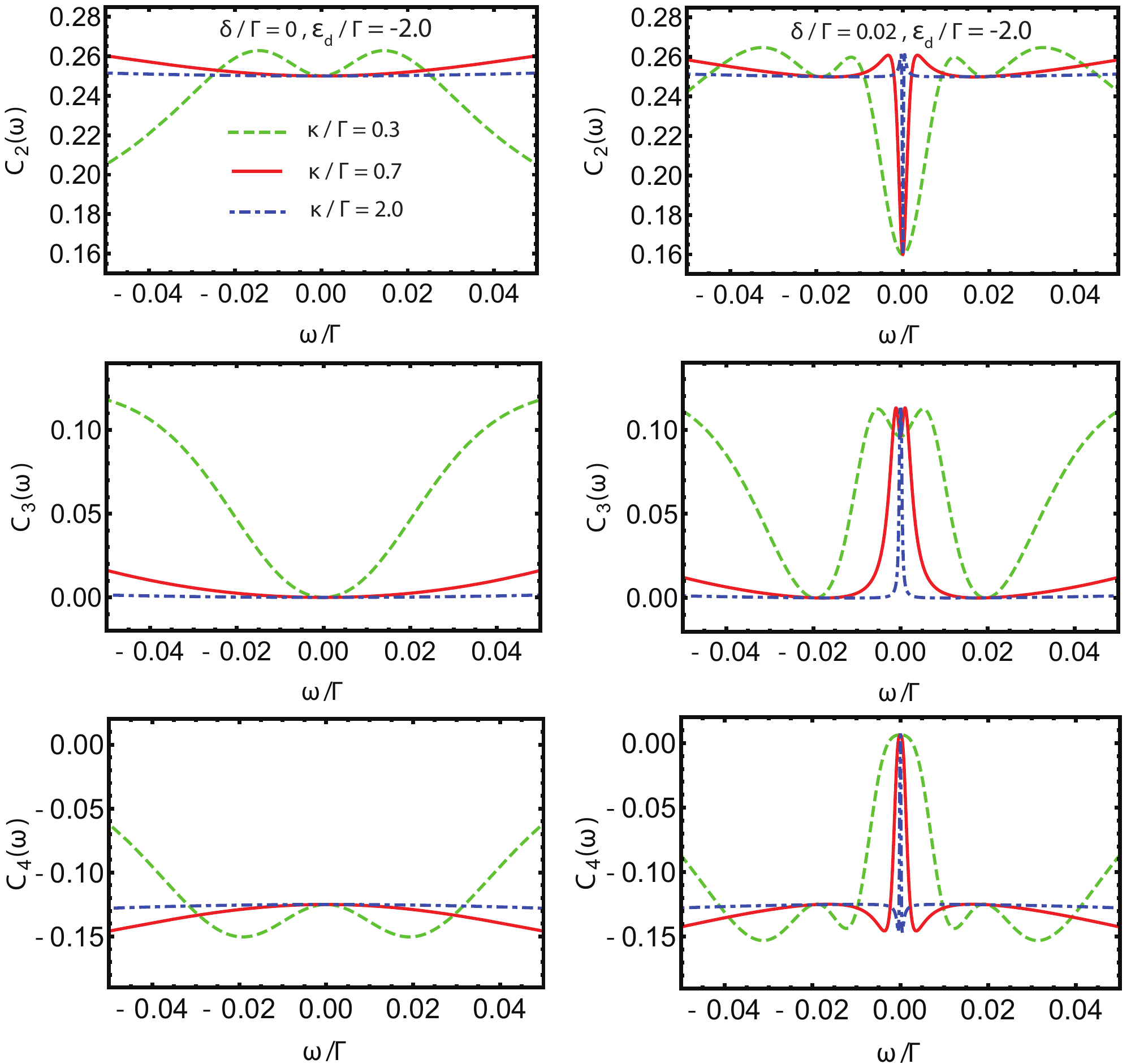}
 \caption{ The cumulant spectrum $C_n(\omega)$ for $n=2,3,4$
for different $\kappa$, $\delta=0$ (left panel) and $\delta/\Gamma=0.02$ (right panel).
Here we set $\epsilon_d/\Gamma=-2.0$, $\Gamma_L=\Gamma_R$.}
\label{fig:CUMspectrum}
\end{figure*}

We now discuss higher order cumulants $n>1$. One can check that the expressions for the shot noise (as well as other higher order cumulants)
through the left and right leads are the same in the $eV\rightarrow 0$ limit. Beyond $eV\rightarrow 0$ limit, this relation only
holds for the symmetric couplings $\Gamma_L=\Gamma_R$.
Therefore, we set $\lambda_R=0$ from now on and study current fluctuations through the left junction only. By  expanding the CGF in terms of $e^{i\lambda_L}$, one finds
\begin{equation}
 \chi (\lambda_L) = \Big(\frac{\Gamma_R}{\Gamma_R+\Gamma_L}\Big)^M \sum_{n=0}^{\infty} {M \choose n} \Big(\frac{\Gamma_L}{\Gamma_R}\Big)^n e^{i n \lambda_L}.
\end{equation}
The probability $\mathcal{P}_q$ can be obtained by the Fourier transform
\begin{eqnarray}
 \mathcal{P}_q = \frac{1}{2\pi}\int_{0}^{2\pi}d\lambda_L e^{-i q \lambda_L} \chi(\lambda_L)\nonumber\\={M \choose q} \Big(\frac{\Gamma_R}{\Gamma_R+\Gamma_L}\Big)^M \Big(\frac{\Gamma_L}{\Gamma_R}\Big)^q.
\end{eqnarray}
As expected, the generating function in the presence of MZM coupling $\kappa\neq 0$ is still described by the binomial distribution. However, the cumulants, defined as
\begin{equation}
 C_n(0)=\frac{\langle\langle \delta^n q \rangle\rangle}{M} = (-i)^n \frac{1}{M}
 \frac{\partial^{n}}{\partial\lambda_L^n} \rm{ln} \chi(\lambda) \Big|_{\lambda_{L,R}=0},
\end{equation}
follow a peculiar pattern at $\Gamma_L=\Gamma_R$
\begin{eqnarray}
 &&\left\{C_1(0)\cdots C_8(0) \right\} =
\left \{\frac{1}{2}, \frac{1}{4},0,-\frac{1}{8},0,\frac{1}{4},0,-\frac{17}{16}\right\}, \nonumber\\
  &&\quad\quad C_n(0)=\frac{E_{n-1}(1)}{2},
  \label{eq:cumulants_MZM}
\end{eqnarray}
with $E_{n}(x)$ being the Euler polynomial. Contrary to the case without MZM, higher order cumulants are nonzero at $\Gamma_L=\Gamma_R$ and are independent of $\epsilon_d$ and $\kappa$.

The dependence of the cumulants on Majorana splitting energy $\delta$ and finite voltage bias (beyond linear in $V$ contributions) can be obtained using Eq. (\ref{eq:CGF_omega}). In this case, one needs to perform an integration of the cumulant spectrum $C_{n}(\omega)$ over $\omega\in (-eV/2,eV/2)$, where
\begin{align}
 C_{n}(\omega) =  \frac{\langle\langle \delta^n q \rangle\rangle(\omega)}{M}=
     \left.  (-i)^n \frac{1}{2} \frac{\partial^{n}}{\partial\lambda_L^n} \ln \left(1+\frac{\mathcal{N}(\omega)}{\mathcal{D}(\omega)} \right)  \right|_{\lambda_{L,R}=0}.
\end{align}
The frequency dependence of the cumulants for $\epsilon_d/\Gamma=-2$ with different $\kappa$ and $\delta$ are shown in Fig. \ref{fig:CUMspectrum}
(left panel: $\delta=0$; right panel: $\delta/\Gamma=0.02$).  As one can see, the cumulants exhibit plateaus corresponding to the universal values, see Eq.\eqref{eq:cumulants_MZM}, in the frequency range $\omega < {\rm min} \{ \Gamma, \kappa^2/\Gamma \}$ which allows one to distinguish the Majorana physics from the other non-Majorana effects.

Next,we consider the effect of finite Majorana degeneracy splitting $\delta\neq 0$ which affects
the cumulant spectrum $C_n(\omega)$ at small frequencies $\omega\rightarrow 0$. One can see that, provided $\kappa\gtrsim \Gamma$,
the cummulants with $\delta=0$ and $\delta\neq 0$ are similar for $|\omega|\gg\delta$. Therefore, in order to
observe the universal values of $C_n$s, one has to adjust the source-drain bias $V$ to the appropriate
regime: ${\rm min}\{\kappa^2/\Gamma,\Gamma\}\gg eV\gg\delta$. One can notice that there is also
a redistribution of the spectral weight for small $\kappa/\Gamma$. Therefore, large $\kappa/\Gamma \gtrsim 1$ limit
is more favorable for the experimental measurement of the cumulants.

\begin{figure*}
 \centering
\includegraphics[width=5.0in,clip]{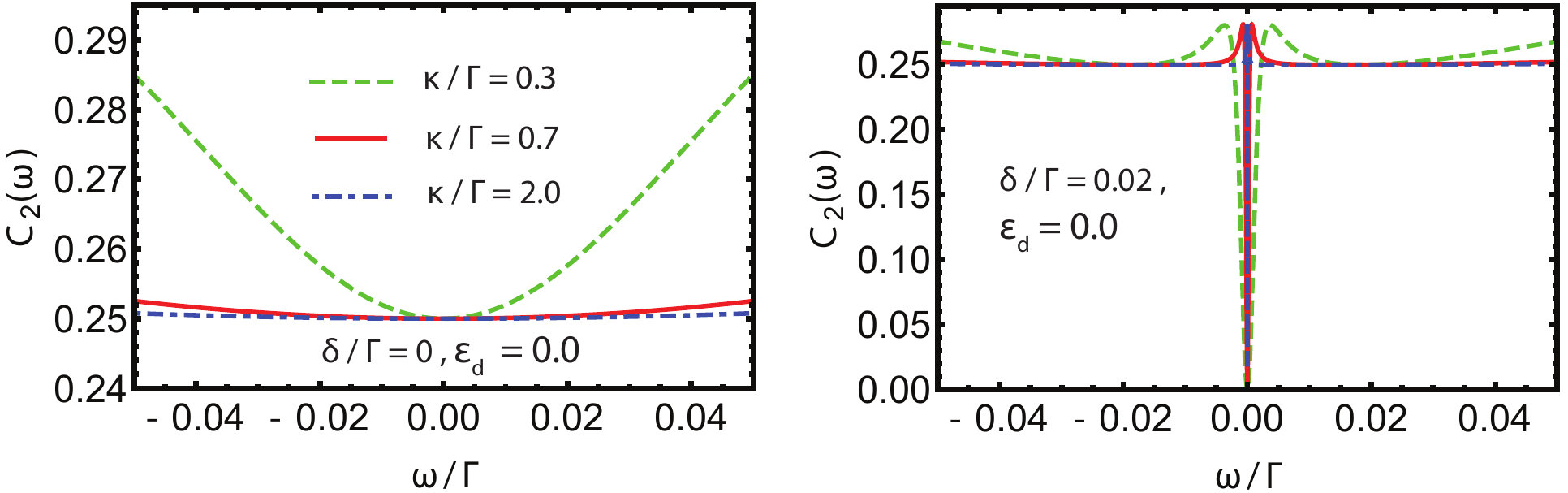}
 \caption{ The $\epsilon_d=0.0$ result of the second cumulant spectrum $C_2(\omega)$
for different $\kappa$, $\delta=0$ (left panel) and $\delta/\Gamma=0.02$ (right panel).
Here we take $\Gamma_L=\Gamma_R$.}
\label{fig:CUMspectrumEd0}
\end{figure*}

Finally, we plot the second cumulant spectrum $C_2(\omega)$ for $\epsilon_d=0$ in Fig. \ref{fig:CUMspectrumEd0}.
We can see that although the quantitative value show small changes compared to $\epsilon_d/\Gamma=-2$ result, the
conditions for the source-drain bias shown above still hold indicating that our results are robust against changes of $\epsilon_d$.

\section{Weakly interacting spinful QD coupled to a MZM}\label{sec:Perturbation}


\begin{figure*}[htp]
\centering
\includegraphics[width=6.5in,clip]{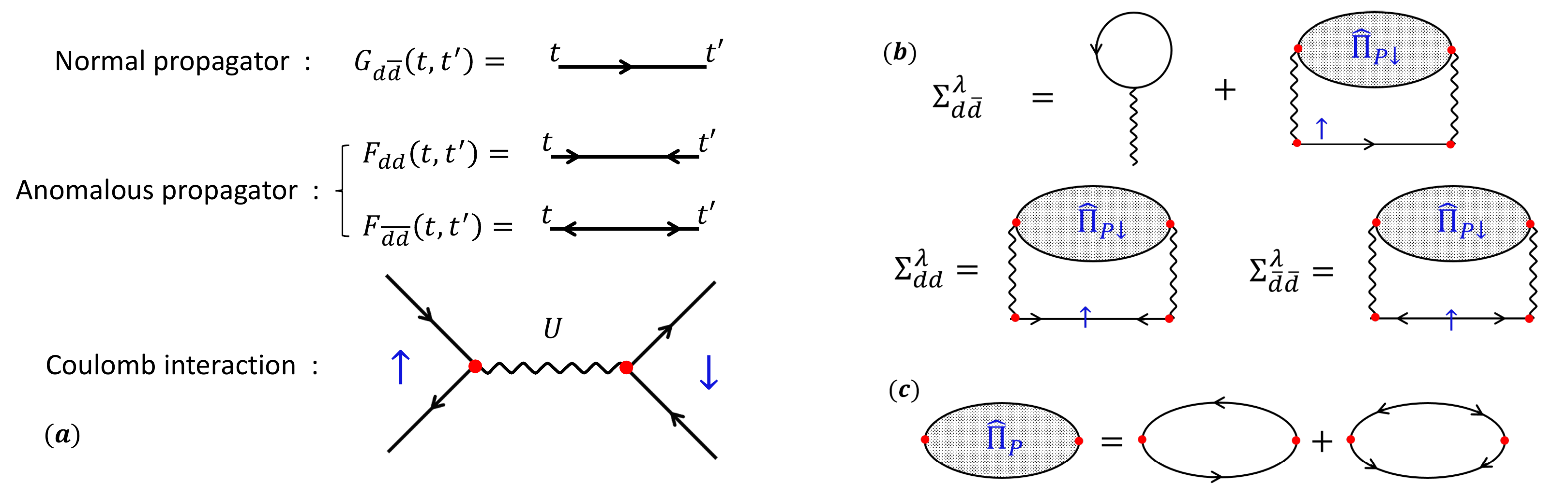}
\caption{a) Diagrammatic representation of the normal and anomalous impurity Green functions and the Coulomb interaction;
b) and c) diagrammatic representation of the self-energy due to leading order corrections of the Coulomb interaction.}
\label{fig:DiagramDef}
\end{figure*}

In this section, we consider the spinful model for a QD coupled to a MZM which is relevant in the context of the Majorana proposals involving topological insulators~\cite{Fu&Kane08, Fu&Kane09, MajoranaTInanowires}. Indeed, MZM can be localized, for example, at the domain wall between an $s$-wave superconductor and a magnetic insulator of a Quantum Spin Hall insulator heterostructure. Assuming that the magnetic insulator, polarized along $z$-axis, does not affect the spin in QD (i.e. Zeeman splitting in QD is negligibly small), one arrives at the following effective Hamiltonian:
\begin{align}
 H_{\rm QD-MZM}&= \sum_{\sigma}\epsilon_d d_{\sigma}^{\dagger}d_{\sigma}+
     U d_{\uparrow}^{\dagger}d_{\uparrow} d_{\downarrow}^{\dagger}d_{\downarrow}\nonumber\\
     &+i\kappa (d_{\uparrow}+d_{\uparrow}^{\dagger})\gamma_1+ i\delta\gamma_1 \gamma_2.
     \label{eq:HQD_interaction1}
\end{align}
Here we also include the effect of inter-particle interaction $U$ assuming that it is weak, i.e. $U\ll \Gamma,\kappa$. The opposite limit of strong Coulomb interaction in the dot is considered in Sec.\ref{sec:Kondo}.

As shown in the previous section, the CGF for spinless QD becomes universal (i.e. independent of $\epsilon_d$ and $\kappa$) due to the MZM coupling. In the spinful case, only one channel (e.g. spin-up) effectively couples to the MZM, see Eq.\eqref{eq:HQD_interaction1}. Thus, the CGF will also have a nonuniversal contribution from the spin-down channel which is decoupled from MZM. However, as follows from Eq.\eqref{eq:cumulants_NoMZM}, higher-order cumulants (i.e. $n>1$) from the spin-down channel vanish at $\epsilon_d=0$ and $\Gamma_L=\Gamma_R$ enabling one to observe the universal part originating from the spin-up part. Thus, some fine-tuning is 
necessary in this case (as opposed to the strongly interacting case in Sec.\ref{sec:Kondo}). In addition to the aforementioned corrections to the universal features in FCS, one should also consider the effect of Coulomb interactions. Without loss of generality, we set $\lambda_R=0$ and calculate effect interactions on charge fluctuations through the left lead. Our conclusions also apply to charge fluctuations through the right lead.

We now consider effect of weak interactions $U\ll \{\Gamma,\kappa\}$ on FCS. We first calculate the contribution of Coulomb interaction to the self-energy and then obtain the corrections to the CGF in powers of $U$. Up to the second order in $U$,
the corresponding Feynman diagrams are shown in Fig. \ref{fig:DiagramDef}. The linear in $U$ contribution to the self-energy $\Sigma_{d\bar{d}}^{\lambda_L}$, see Fig. \ref{fig:DiagramDef}, merely represents the renormalization of the QD energy level $\epsilon_d$. This is a trivial interaction effect which does not modify our previous conclusions. We, therefore, focus on $U^2$ contributions. The self-energy in the Nambu space has the following form
\begin{equation}
 \Sigma^{\lambda_L}=
 \begin{pmatrix}
    \Sigma_{d\bar{d}}^{\lambda_L} & \Sigma_{dd}^{\lambda_L} \\
    \Sigma_{\bar{d}\bar{d}}^{\lambda_L} & \Sigma_{\bar{d}d}^{\lambda_L}
 \end{pmatrix},
\end{equation}
where $\Sigma_{\bar{d}d}^{\lambda_L}$ and $ \Sigma_{\bar{d}\bar{d}}^{\lambda_L}$ are particle-hole conjugation
of $\Sigma_{d\bar{d}}^{\lambda_L}$ and $\Sigma_{dd}^{\lambda_L}$. We note that all the Green's functions here depend on the counting field $\lambda_L$. The details of the calculation of $\Sigma^{\lambda_L}$ is presented in the Appendix~\ref{app:Correction}. After some manipulations, the cumulant generating function for each spin-channel can be written as
\begin{align}
 \ln \chi (\lambda) &= \frac{\mathcal{T}}{2} \int_{-\infty}^{\infty} \frac{d \omega}{2\pi} \ln
 \frac{\det\Big[ \big[\breve{Q}_{dd,U=0}^{\lambda_L}\big]^{-1} - \Sigma^{\lambda_L} \Big]}
               {\det\Big[  \big[\breve{Q}_{dd,U=0}^{\lambda_L=0}\big]^{-1} - \Sigma^{\lambda_L=0}  \Big]}\nonumber\\
            &= \frac{\mathcal{T}}{2} \int_{-\infty}^{\infty} \frac{d \omega}{2\pi} \ln
               \frac{\det\Big[ \big[\breve{Q}_{dd,U=0}^{\lambda_L}\big]^{-1} \Big]}
               {\det\Big[  \big[\breve{Q}_{dd,U=0}^{\lambda_L=0}\big]^{-1} \Big]} \label{eq:interaction_corr} \\
             & +\frac{\mathcal{T}}{2} \int_{-\infty}^{\infty} \frac{d \omega}{2\pi} \ln
               \frac{\det\Big[  \mathbb{I}_{4\times 4}- \breve{Q}_{dd,U=0}^{\lambda_L} \Sigma^{\lambda_L} \Big]}
               {\det\Big[  \mathbb{I}_{4\times 4}- \breve{Q}_{dd,U=0}^{\lambda_L=0} \Sigma^{\lambda_L=0} \Big]},\nonumber
\end{align}
where
\begin{eqnarray}
 \big[\breve{Q}_{dd,U=0}^{\lambda_L}\big]^{-1} &=&  \big[\breve{Q}_{0,dd}^{\lambda_L}\big]^{-1}
             -\sum_{\alpha} \big( \sum_{i} M_{T,\alpha}^i\otimes \gamma^i \big)^{\dagger} \breve{Q}_{0,\alpha}  \nonumber\\
            &&\quad\quad\quad\quad \quad   \times \big( \sum_{i} M_{T,\alpha}^i\otimes \gamma^i \big),
\end{eqnarray}
The first term in the generating function corresponds to the result for noninteracting case, see Eq. (\ref{eq:generatingU0}) whereas the second term originates from the interaction-induced corrections. We consider weak Coulomb interactions $U\ll \{\Gamma,\lambda,eV\}$, and keep the leading order terms in $U$: $\breve{Q}_{dd,U=0}^{\lambda_L} \Sigma^{\lambda_L}\sim U^2 / \mathrm{min}\{\Gamma,\lambda,eV\}^2\ll 1$. By expanding the second term in Eq.\eqref{eq:interaction_corr} up to quadratic order in $U$, we obtain
\begin{eqnarray}
 \ln \chi (\lambda) &\approx& \frac{\mathcal{T}}{2} \int_{-\infty}^{\infty} \frac{d \omega}{2\pi} \ln
               \frac{\det\Big[ \big[\breve{Q}_{dd,U=0}^{\lambda_L}\big]^{-1} \Big]}
               {\det\Big[  \big[\breve{Q}_{dd,U=0}^{\lambda_L=0}\big]^{-1} \Big]} \nonumber\\
              && - \frac{\mathcal{T}}{2} \int_{-\infty}^{\infty} \frac{d \omega}{2\pi}
               \Big[ \rm{Tr}\big( \breve{Q}_{dd,U=0}^{\lambda_L} \Sigma^{\lambda_L} \big) \nonumber\\
              && \quad\quad - \rm{Tr}\big( \breve{Q}_{dd,U=0}^{\lambda_L=0} \Sigma^{\lambda_L=0} \big) \Big],
              \label{eq:C_interaction}
\end{eqnarray}
where we used the relation $\det(\mathbb{I}+x\breve{A})\approx  1+ x\rm{Tr}(\breve{A})$ for $x\ll 1$.

\begin{figure*}
\centering
\vspace{0.2in}
\includegraphics[width=6.0in,clip]{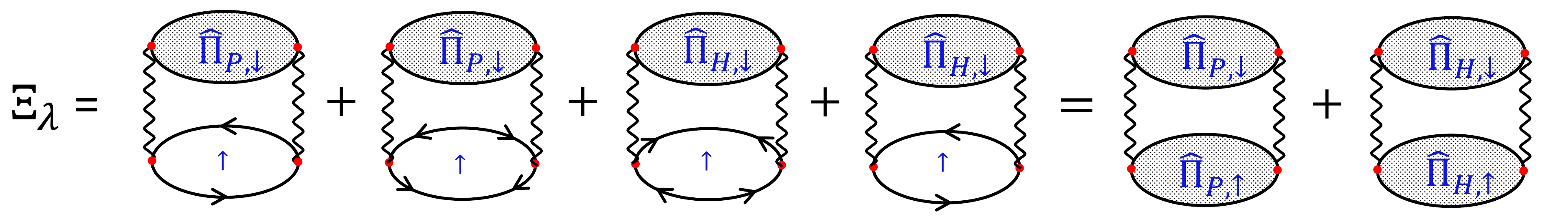}
\caption{Diagrammatic representation of the cumulant generating function up to the leading order corrections of the Coulomb interaction.
}
\label{fig:CGFUDiag22}
\end{figure*}


After some manipulations (see Appendix~\ref{app:Correction}), the cumulant generating function for small $U$ can be written as
\begin{equation}
 \ln \chi_{\sigma}(\lambda_L,U) \approx  \ln \chi_{\sigma}(\lambda_L,U=0)-\frac{\mathcal{T} U^2}{2} \Xi_{\lambda}^{\widetilde{K}}.
 \label{eq:C_interaction_Formula}
\end{equation}
Note that the matrix formalism of the Green functions has the form
$\bigl(\begin{smallmatrix}
G^{R}&G^{K}\\ G^{\widetilde{K}}&G^{A}
\end{smallmatrix} \bigr)$ (i.e. with a L-O rotation). However, for the convenience of the calculation,
we will also use the matrix Green functions in the Schwinger - Keldysh space (without L-O rotation).
Here the function $\Xi_{\lambda}$ ($\lambda$ here means $(\lambda_L,\lambda_R)$) in the Schwinger - Keldysh space has the form
\begin{equation}
\Xi_{\lambda} =
 \begin{pmatrix}
  \Xi_{\lambda}^{T} & \Xi_{\lambda}^{<} \\
  \Xi_{\lambda}^{>} & \Xi_{\lambda}^{\widetilde{T}}
 \end{pmatrix},
\end{equation}
where
\begin{align}
\Xi_{\lambda}^{T} &= \int_{-\infty}^{\infty} \frac{d\Omega}{2\pi} \Big( \hat{\Pi}_{P,\uparrow}^{T}(-\Omega)\hat{\Pi}_{P,\downarrow}^{T}(\Omega)
                        +\hat{\Pi}_{H,\uparrow}^{T}(-\Omega)\hat{\Pi}_{H,\downarrow}^{T}(\Omega) \Big),\nonumber\\
\Xi_{\lambda}^{\widetilde{T}} &= \int_{-\infty}^{\infty} \frac{d\Omega}{2\pi} \Big( \hat{\Pi}_{P,\uparrow}^{\widetilde{T}}(-\Omega)\hat{\Pi}_{P,\downarrow}^{\widetilde{T}}(\Omega)
                        +\hat{\Pi}_{H,\uparrow}^{\widetilde{T}}(-\Omega)\hat{\Pi}_{H,\downarrow}^{\widetilde{T}}(\Omega) \Big),\nonumber\\
\Xi_{\lambda}^{<} &= \int_{-\infty}^{\infty} \frac{d\Omega}{2\pi} \Big( \hat{\Pi}_{P,\uparrow}^{<}(-\Omega)\hat{\Pi}_{P,\downarrow}^{<}(\Omega)
                        +\hat{\Pi}_{H,\uparrow}^{<}(-\Omega)\hat{\Pi}_{H,\downarrow}^{<}(\Omega) \Big),\nonumber\\
\Xi_{\lambda}^{>} &= \int_{-\infty}^{\infty} \frac{d\Omega}{2\pi} \Big( \hat{\Pi}_{P,\uparrow}^{>}(-\Omega)\hat{\Pi}_{P,\downarrow}^{>}(\Omega)
                        +\hat{\Pi}_{H,\uparrow}^{>}(-\Omega)\hat{\Pi}_{H,\downarrow}^{>}(\Omega) \Big),
\end{align}
and
\begin{equation}
 \Xi_{\lambda}^{\widetilde{K}}=\Big(\Xi_{\lambda}^{T}+\Xi_{\lambda}^{\widetilde{T}}-\Xi_{\lambda}^{<}-\Xi_{\lambda}^{>}\Big)/2.
\end{equation}
The polarization functions $\hat{\Pi}_{P,\sigma}^{\alpha}$ and $\hat{\Pi}_{H,\sigma}^{\alpha}$ (see Fig. \ref{fig:DiagramDef})
are calculated in Appendix \ref{app:Correction}.
One can notice that, for $\epsilon_d=0$, the particle parts are exactly the same as the hole parts: $\hat{\Pi}_{P,\sigma}=\hat{\Pi}_{H,\sigma}$.
Thus, the functions $\Xi_{\lambda}$ for spin-up channel have the same form as those for spin-down channel, and we drop the spin-index in  $\Xi_{\lambda}$ from now on.
Lastly, it is well-known that the function $\Xi_{\lambda=0}^{\widetilde{K}}=0$ (here $\lambda=0$ means $\lambda_L=\lambda_R=0$) vanishes because of causality and unitarity. However, the presence of artificial counting field $\lambda(t)$ having different sign on forward branch and backward branches of the Keldysh contour breaks unitarity. Therefore, the relation $\Xi_{\lambda=0}^{\widetilde{K}}=0$ does not hold for $\lambda\neq 0$, and we have to evaluate it explicitly.
In general, the calculation of $\Xi_{\lambda,A}^{\widetilde{K}}$ is not very illuminating but some simplification can be obtained by expanding the interaction-induced corrections in powers of $eV$ assuming that $eV\rightarrow 0$. After
some manupilations (see Appendices \ref{app:Expansion}, \ref{app:Derivation_XiA}, and \ref{app:Derivation_XiB}),
we find that the leading contribution to CGF is proportional to $V^3$:
\begin{equation}
 \Xi_{\lambda,A}^{\widetilde{K}} = (eV)^3 \Big( \Xi_{\lambda,A}^{\widetilde{K},(3)}+
         \Xi_{\lambda,B}^{\widetilde{K},(3)}\Big),
\end{equation}
where the functions $\Xi_{\lambda,A}^{\widetilde{K},(3)}$ and $\Xi_{\lambda,B}^{\widetilde{K},(3)}$ are given by
\begin{widetext}
\begin{eqnarray}
 \Xi_{\lambda,A}^{\widetilde{K},(3)} &=& \frac{ (e^{-i\lambda_L}-1)}{6\pi \Gamma^4} \mathbf{O}(\widetilde{\kappa})
       + (2-\frac{\pi^2}{4}) \frac{(e^{i\lambda_L}-1) \big[(e^{i\lambda_L}-1)-2(e^{i\lambda_L}+3)\widetilde{\kappa}^2 \big]}
       {12\pi^3 (e^{i\lambda_L}+1)^2\Gamma^4 \widetilde{\kappa}^2},\label{eq:XiA}\\
\Xi_{\lambda,B}^{\widetilde{K},(3)} &=& \frac{1}{24\pi^3\Gamma^4\kappa^2\big(1+e^{i\lambda_L}\big)^2}
             \Big[ 15\kappa^2\big(e^{i\lambda_L}-1\big) + \kappa^2 \big(e^{2i\lambda_L}-1\big)
              +(-2\Gamma^2-15\kappa^2) \big(e^{-i\lambda_L}-1\big) \nonumber\\
              &&+(\Gamma^2-8\kappa^2)\big(e^{-2i\lambda_L}-1\big)\Big]
              +\frac{\hat{\Pi}_{P2,\uparrow}^{T}(0^+,0)-\hat{\Pi}_{P2,\uparrow}^{\widetilde{T}}(0^+,0)}{6\pi^2\Gamma^3}
              \big(e^{-i\lambda_L}-1\big),\label{eq:XiB}
\end{eqnarray}
where
\begin{equation}
 \mathbf{O}(\widetilde{\kappa}) = \int_{-\infty}^{\infty} \frac{d \widetilde{\Omega}}{2\pi} \Big[
             \hat{\Pi}_{P,\uparrow}^{T}(-\widetilde{\Omega},0) \frac{-i}{(|\widetilde{\Omega}|+i)^2}
             + \hat{\Pi}_{P,\uparrow}^{\widetilde{T}}(-\widetilde{\Omega},0) \frac{-i}{(|\widetilde{\Omega}|-i)^2}\Big],
              \text{with $\widetilde{\Omega}=\frac{\Omega}{\Gamma}$, $\widetilde{\kappa}=\frac{\kappa}{\Gamma}$.}
\end{equation}
One can check that indeed above expression vanish for $\lambda_L=0$, as required by unitarity. The integral in Eq. (\ref{eq:XiA}) is a dimensionless number depending only on $\kappa/\Gamma$. The functions $\hat{\Pi}_{P2,\uparrow}^{T,\widetilde{T}}(0^+,0)$ in Eqs. \eqref{eq:XiB} are defined in Eq.(\ref{eq:Pi2_Def});
these functions depends on $\Gamma$ and $\kappa$. \cite{Abanov&Ivanov09,BelzigPRB05,FlindtPRB10,Utsumi13}
\footnote{We note here that Eqs. \eqref{eq:XiA} and \eqref{eq:XiB} contain terms proportional
to $(1+e^{i\lambda_L})^{-2}$ which can be represented as an infinite series in $e^{i\lambda_L}$.
Similar terms in CGF also appear in other interacting systems~\cite{Abanov&Ivanov09,BelzigPRB05,FlindtPRB10,Utsumi13},
and there has been some debate as to their interpretation. We note, however, that physical observables (cumulants)
are well-defined even in the presence of these terms, cf Eq.\eqref{eq:corrU1}.}

Combining all the terms, the final expression for the cumulant generating function of the spinful QD at $O(U^2)$-level can be written as
\begin{align}\label{eq:cumulants_int}
 \ln \chi(\lambda_L,U) \approx  \ln \chi_{\uparrow}(\lambda_L,U=0)+\ln \chi_{\downarrow}(\lambda_L,U=0)
 - \mathcal{T} U^2 (eV)^3 \Big( \Xi_{\lambda,A}^{\widetilde{K},(3)}
  + \Xi_{\lambda,B}^{\widetilde{K},(3)}\Big).
\end{align}
and the leading correction for the current and shot noise (for left junction current) can be written as
\begin{align}\label{eq:corrU1}
 \delta I_L^U &= \frac{e}{h}\frac{U^2 (eV)^3}{\Gamma^4} \Bigg( \frac{ \mathbf{O}(\widetilde{\kappa})}{6 \pi} + \left(2-\frac{\pi^2}{4}\right) \frac{1}{6 \pi^3 } - \frac{1}{2\pi^3 }
              +\frac{\hat{\Pi}_{P2,\uparrow}^{T}(0^+,0)-\hat{\Pi}_{P2,\uparrow}^{\widetilde{T}}(0^+,0)}{6\pi^2} \Bigg),\\
 \delta S_{LL}^U &= - \frac{e}{h} \frac{U^2 (eV)^3}{\Gamma^4} \Bigg( \frac{ \mathbf{O}(\widetilde{\kappa})}{6 \pi } + \left(2-\frac{\pi^2}{4}\right) \frac{1+2\widetilde{\kappa}^2}{24 \pi^3  \widetilde{\kappa}^2}
                 + \frac{1-62\widetilde{\kappa}^2}{48 \pi^3  \widetilde{\kappa}^2}
                 +\frac{\hat{\Pi}_{P2,\uparrow}^{T}(0^+,0)-\hat{\Pi}_{P2,\uparrow}^{\widetilde{T}}(0^+,0)}{6\pi^2} \Bigg).
\end{align}
Thus, the leading order correction to the generating function in the presence of MZM coupling is of the order of $(eV)^3$ which is the same in the case without MZM considered in Ref. \onlinecite{Gogolin&Komnik06}.  Since the leading order correction to the cumulants $C_n$ is of the order of $(eV)^2$, the interaction-induced corrections do not affect cumulants at small bias. Therefore, we expect that one can observe the universal values of the cumulants discussed  in Sec.\ref{sec:NonInteracting} in realistic experimental conditions.
\end{widetext}

\begin{figure*}[htp]
\centering
\vspace{0.2in}
\includegraphics[width=5.0in,clip]{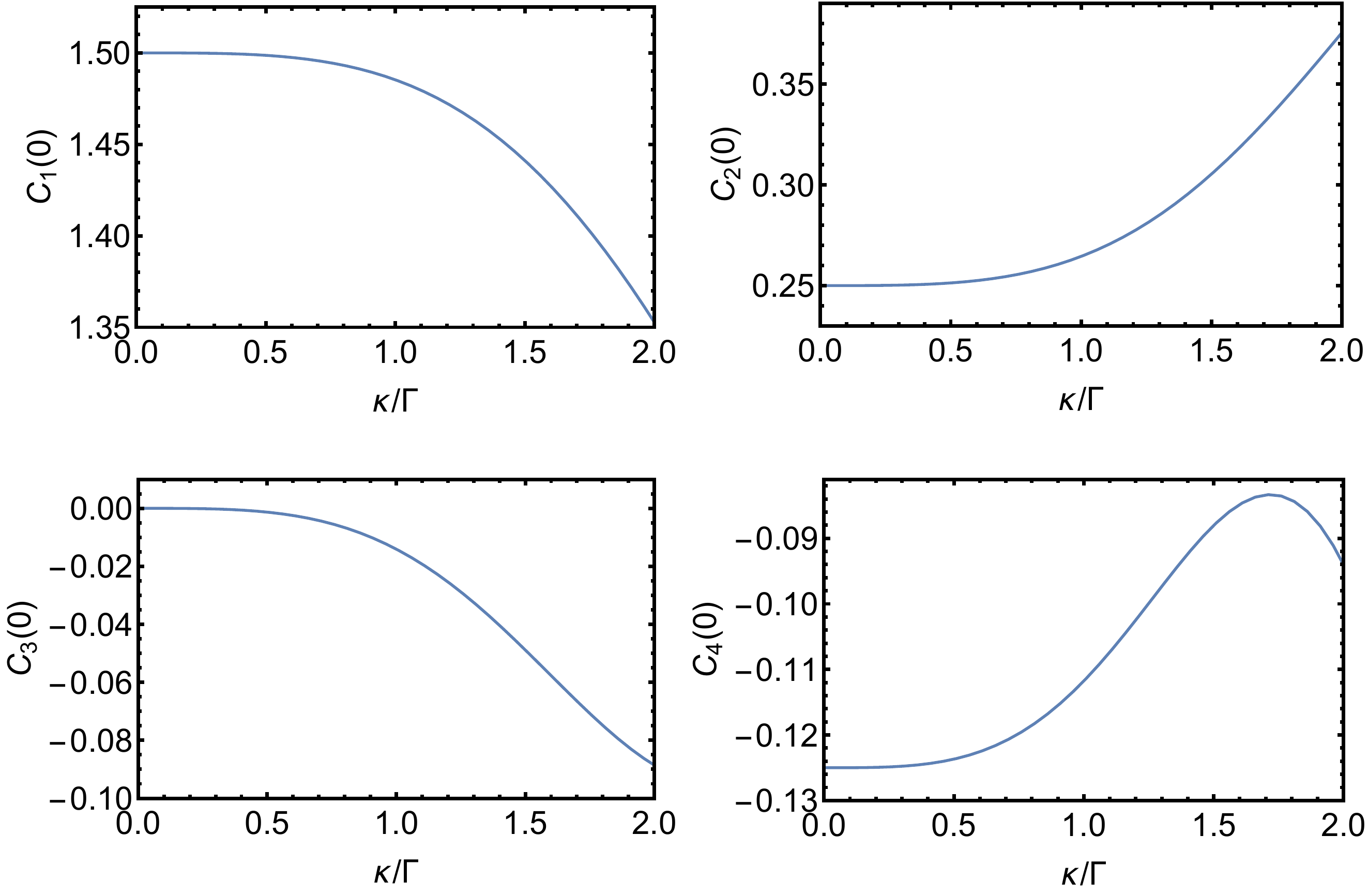}
\caption{Dependence of the cumulants $C_n(0)=\langle\langle \delta^n q\rangle\rangle / M$ ($\delta=0$ and $eV\rightarrow 0$ limit) on the Majorana coupling $\kappa$
in Kondo regime from SBMF calculation.
We choose $\Gamma_L=\Gamma_R$, $\epsilon_d/\Gamma=-10.0$, and band width $\Lambda/\Gamma=30.0$.
}
\label{fig:SBMF_Delta0eV0_CUM}
\vspace{0.2in}
\end{figure*}

\section{Strongly interacting spinful QD with MZM: interplay of Kondo and Majorana couplings}\label{sec:Kondo}

\subsection{Theoretical model}

In this section, we study another nontrivial case corresponding to a strongly interacting spinful QD coupled to a MZM. The Hamiltonian for the system reads
\begin{eqnarray}
 H_{\rm QD-MZM} &=& \sum_{\sigma}\epsilon_d d_{\sigma}^{\dagger}d_{\sigma}+
     U d_{\uparrow}^{\dagger}d_{\uparrow} d_{\downarrow}^{\dagger}d_{\downarrow}\nonumber\\
     && + i\kappa (d_{\uparrow}+d_{\uparrow}^{\dagger})\gamma_1+ i\delta\gamma_1 \gamma_2.
     \label{eq:HQD_interaction}
\end{eqnarray}
Once again, we assume here that Zeeman splitting in a QD is negligibly small (see discussion after Eq.\eqref{eq:HQD_interaction1}). 
For large  on-site Coulomb interaction $U$ and single-electron occupancy, we have to consider interplay of Kondo and Majorana physics~\cite{ChengPRX2014}. 
In the limit of  single-electron occupancy $\{\Gamma, \kappa\}\ll |\epsilon_d|\ll U$, one can study the problem using a slave boson mean field 
(SBMF) approximation originally developed for an infinite-$U$ Anderson model~\cite{ColemanPRB83,LeeRMP06}. This approach allows one to eliminate double occupancy in the QD and significantly simplify the problem.

\begin{figure*}[t]
 \centering
 \begin{tabular}{@{}cc@{}}
  \includegraphics[width=2.8in,clip]{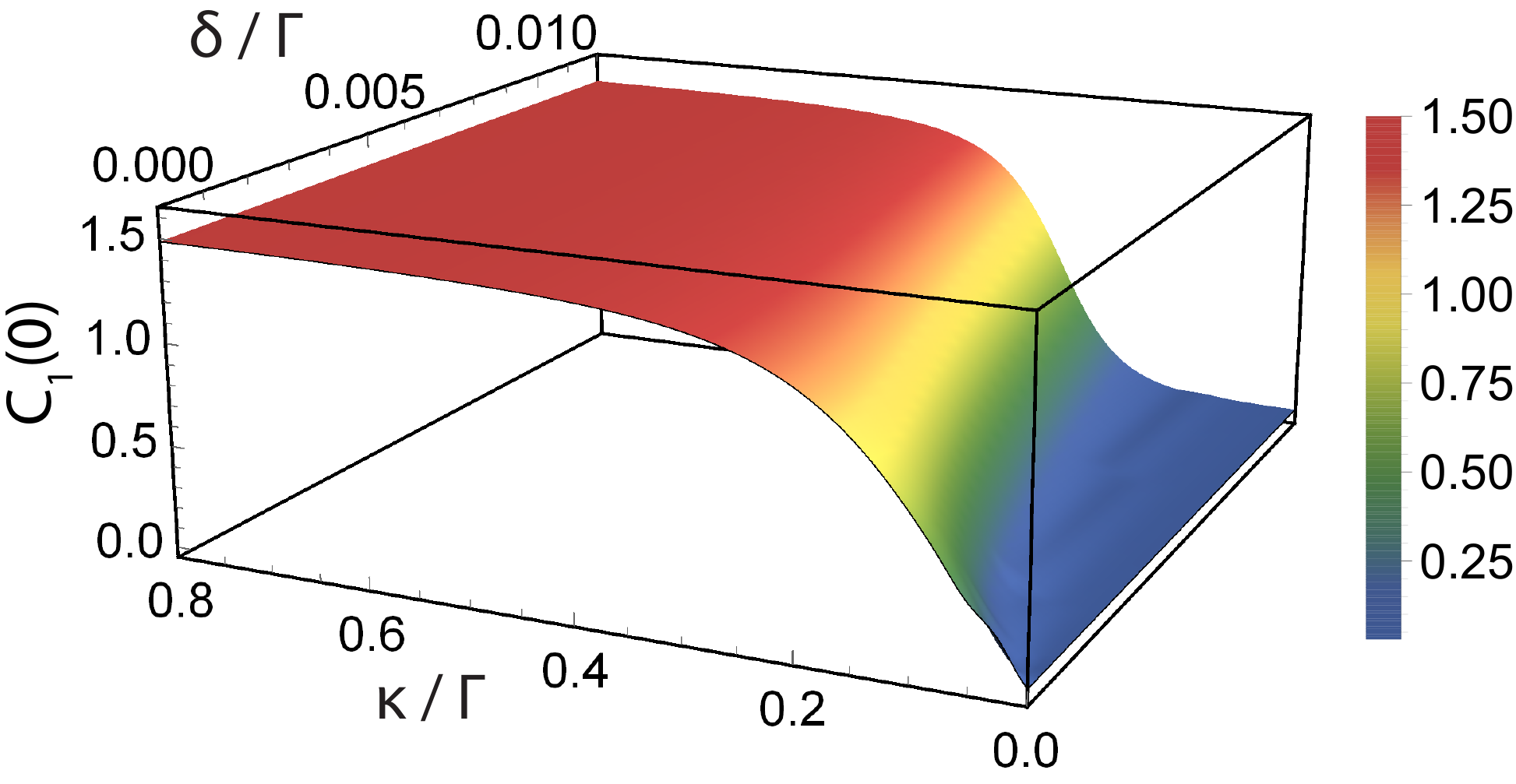} &
  \includegraphics[width=2.8in,clip]{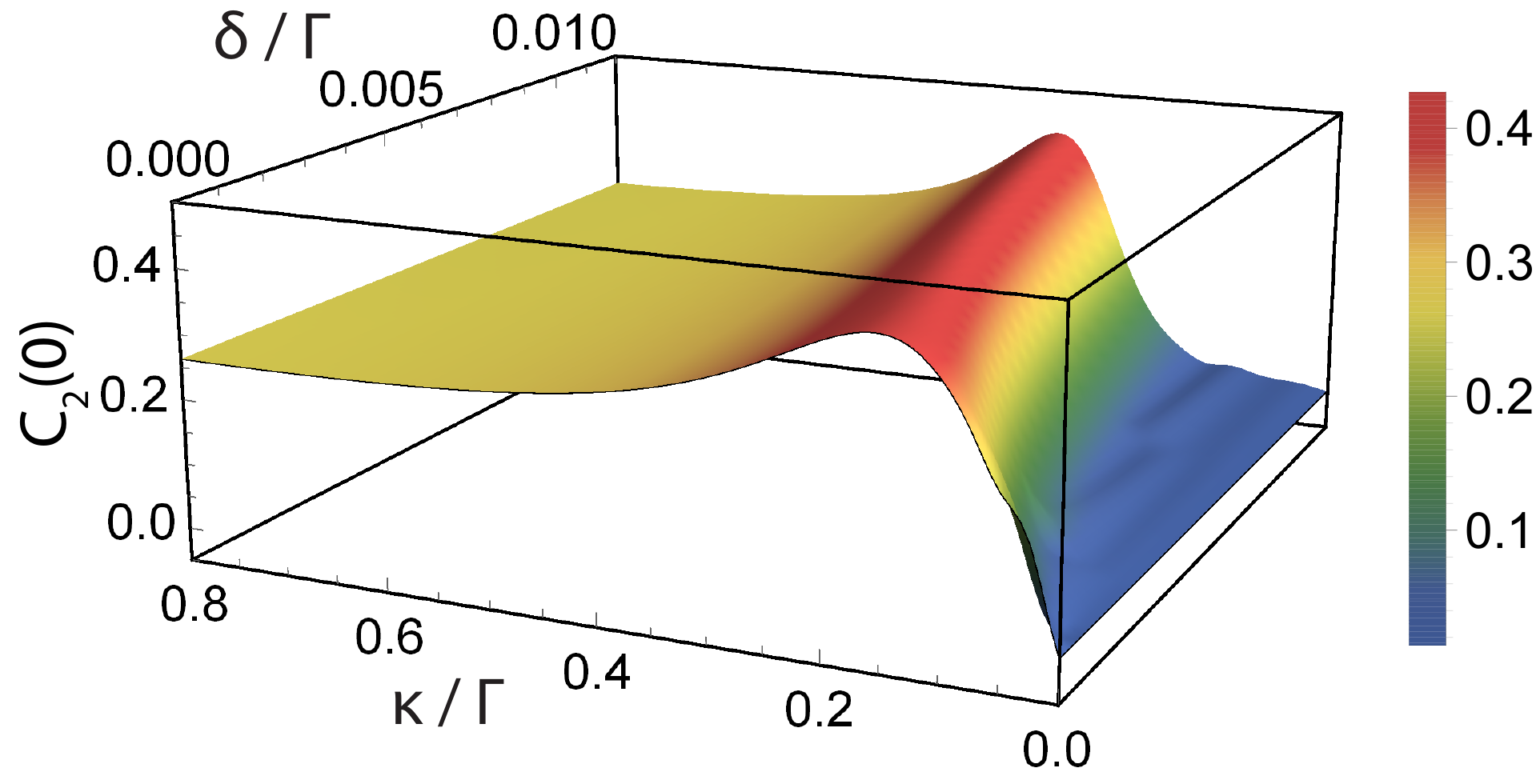} \\
  \includegraphics[width=3.0in,clip]{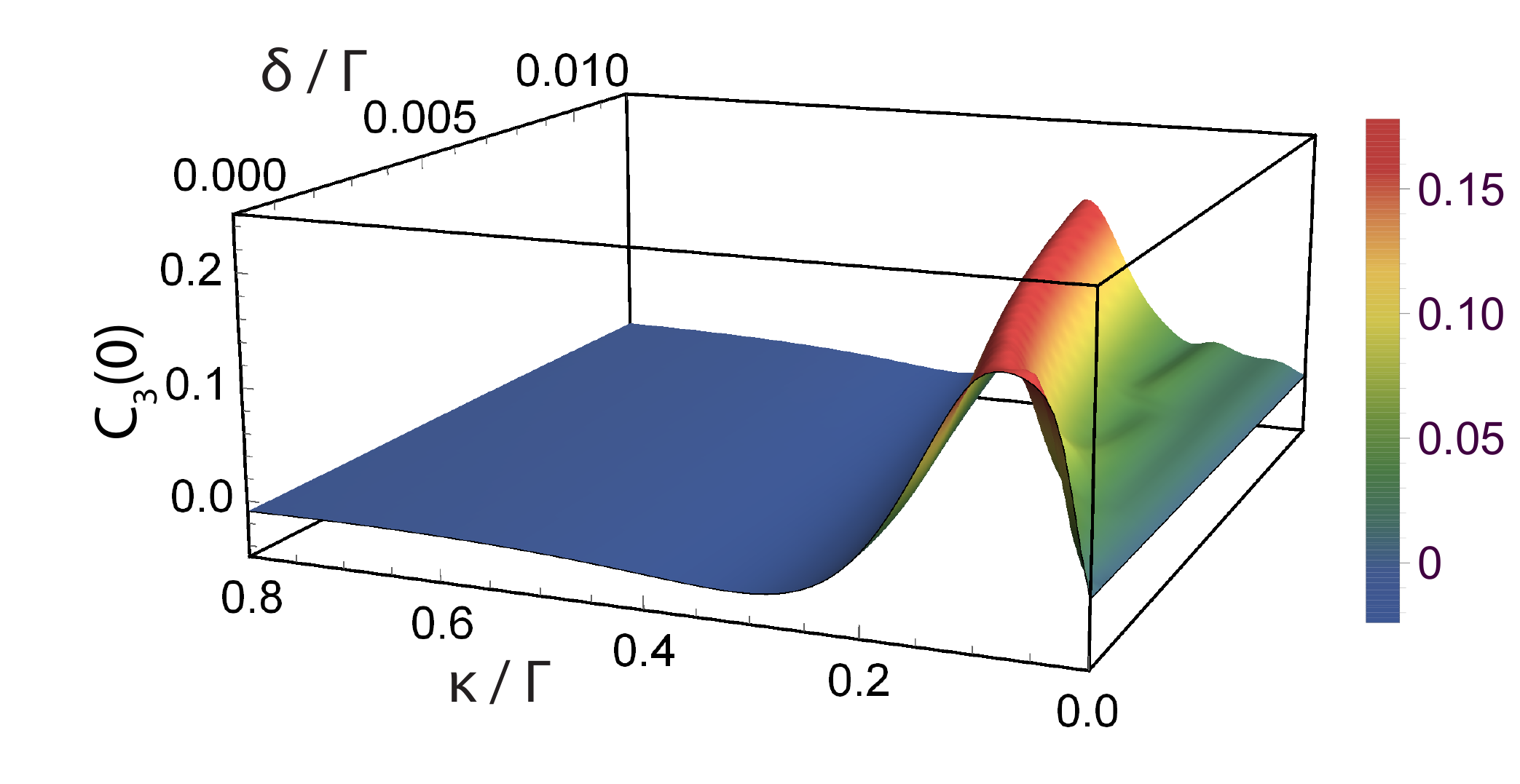} &
  \includegraphics[width=3.0in,clip]{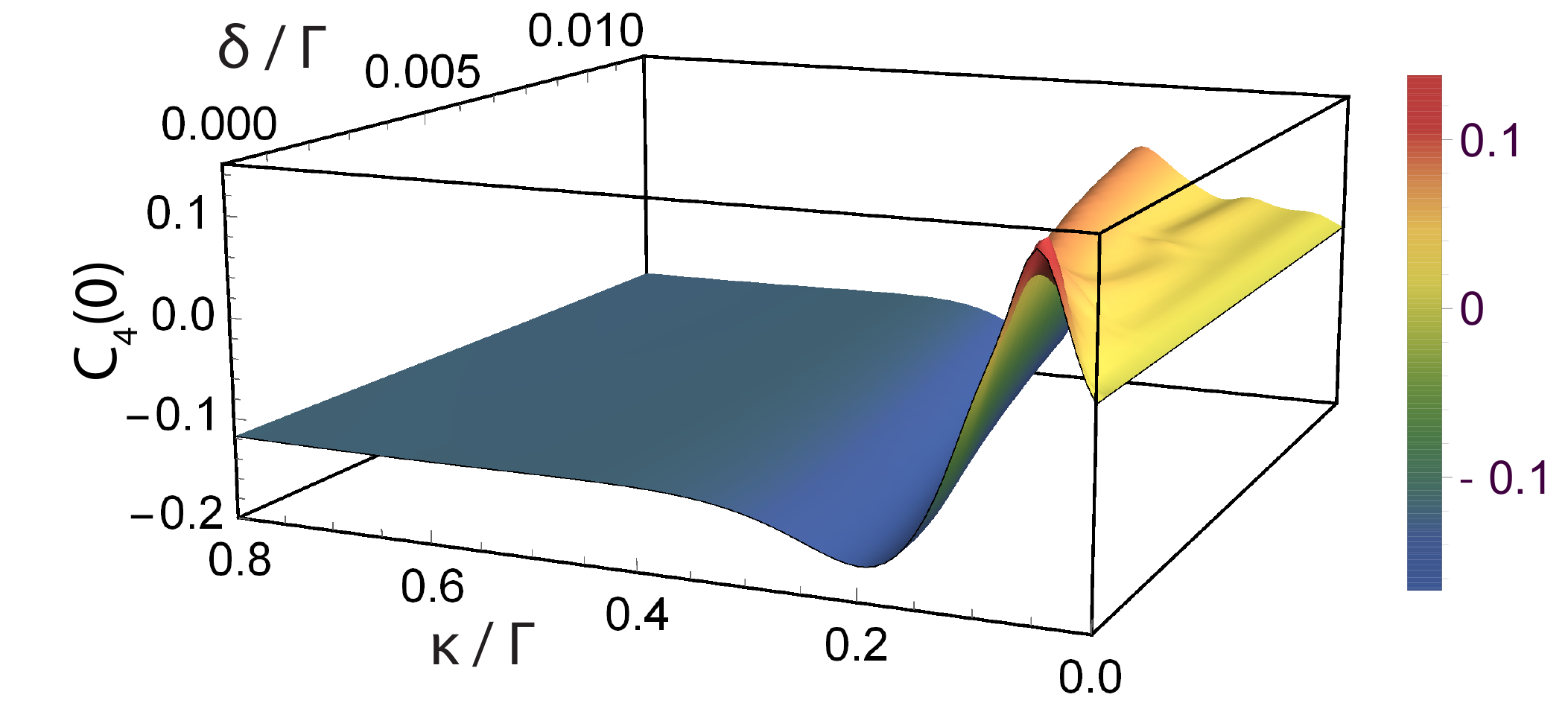}
 \end{tabular}
 \caption{ The cumulants  $C_n(0)$ for $n=1,2,3,4$
as a function $\kappa$ and $\delta$. Here $\epsilon_d/\Gamma=-10.0$, $\Gamma_L=\Gamma_R$, $eV/\Gamma=0.001$, and $\Lambda/\Gamma=30.0$.}
\label{fig:SBMF_CUM}
\end{figure*}

For the sake of completeness, we outline here main steps of SBMF approach. We refer a reader to Ref. \onlinecite{ChengPRX2014} for more details of SBMF calculation in the presence of MZM. We first rewrite fermion operators in QD in terms of the auxiliary boson $b$ and fermion $f_{\sigma}$ operators, i.e. $d_{\sigma}\rightarrow f_{\sigma}b^{\dagger}$. This procedure requires to introduce a constraint $b^{\dagger}b+\sum_{\sigma}f_{\sigma}^{\dagger}f_{\sigma}=1$ on the Hilbert space. After the transformation, effective Hamiltonian becomes
\begin{align}
 &H_{\rm SBMF} = H_{\rm Leads}+ \sum_{\sigma} \epsilon_d f_{\sigma}^{\dagger} f_{\sigma}
                        +i\kappa \gamma_1 (f_{\uparrow}b^{\dagger}+f_{\uparrow}^{\dagger}b) \nonumber\\
              & +\sum_{\alpha=L,R}\sum_{\sigma} t_{\alpha} (\psi_{\sigma,\alpha}^{\dagger}(0) f_{\sigma}b^{\dagger}+h.c.)+ i\delta\gamma_1 \gamma_2
\end{align}
where the lead Hamiltonian $H_{\rm Leads}$ is unchanged. Next, we apply mean field approximation and replace the bosonic operator $b$ and the Lagrangian multiplier $\eta$ enforcing the constraint by their mean-field expectation values. We choose $\langle b \rangle=\langle b^{\dagger} \rangle=b$ to be
a real positive number. The mean field parameter $b$ and $\eta$ can be determined self-consistently by minimizing the free energy \cite{ChengPRX2014}:
\begin{eqnarray}
 && b^2 +\sum_{\sigma} \langle f_{\sigma}^{\dagger} f_{\sigma} \rangle = 1,\\
 && 2b\eta + t \sum_{\alpha=L,R} \sum_{\sigma} (\langle f_{\sigma}^{\dagger}\psi_{\sigma,\alpha}(0) \rangle + c.c.)\nonumber\\
 && +i\kappa \Big\langle \gamma_1 (f_{\uparrow}^{\dagger}+f_{\uparrow}) \Big\rangle = 0.
\end{eqnarray}
Here, we assume the $eV\ll {\rm max} \,\{T_K,\kappa\}$ and thus neglect the dependence on voltage bias $eV$ in the SMBF calculations, see also discussion in Ref.~\cite{FCSSBMF}. Above equations determine thermodynamics of the system, and we are now ready to compute transport properties.
Once the mean field values of the auxiliary parameters are determined in SBMF approximation, spin-up and spin-down channels become decoupled. Thus, effectively the problem reduces to the previous model - spin-up channel is coupled to a MZM whereas the spin-down channel is not. The chemical potential and couplings are renormalized: $\epsilon_d\rightarrow \widetilde{\epsilon}_d=\epsilon_d+\eta$,
$t_{\alpha}\rightarrow \widetilde{t}_{\alpha}=b t_{\alpha}$, $\Gamma_{\alpha}\rightarrow \widetilde{\Gamma}_{\alpha}=b^2 \Gamma_{\alpha}$, and $\kappa\rightarrow \widetilde{\kappa}=b  \kappa$.

Using the results of Sec.\ref{sec:NonInteracting}, it is rather straightforward to obtain the CGF at small voltage bias $eV\rightarrow 0$ in this case
\begin{eqnarray}\label{eq:spinless_cumulant}
  && \frac{ \ln \chi_{\rm KM}}{M}\Bigg|_{eV\rightarrow 0} = \ln\left(\frac{\widetilde{\Gamma}_L e^{i\lambda_L}+\widetilde{\Gamma}_R e^{- i\lambda_R}}{\widetilde{\Gamma}_L+\widetilde{\Gamma}_R}\right)\nonumber\\
      &&\quad\quad+\ln\left( 1+ \frac{4 \widetilde{\Gamma}_L \widetilde{\Gamma}_R}{\widetilde{\epsilon}_d^2+(\widetilde{\Gamma}_L+\widetilde{\Gamma}_R)^2}
      (e^{i(\lambda_L-\lambda_R)}-1) \right).
\end{eqnarray}
Here the first and second terms correspond to the spin-up and spin-down channels, respectively. As discussed below, the renormalized QD energy $\widetilde{\epsilon}_d$
is close to the Fermi level, i.e. $\widetilde{\epsilon}_d \rightarrow 0$, for a large parameter range $\{\kappa,\Gamma\}< |\epsilon_d|$ (so-called universal limit). Thus, in the case of symmetric right-left lead couplings $\Gamma_R=\Gamma_L$, the second term in Eq.~\eqref{eq:spinless_cumulant} is simply given by $i (\lambda_L-\lambda_R)$ and, therefore, does not contribute beyond the first cumulant. Thus, the shot noise as well as other higher order cumulants are once again given by the universal pattern:
\begin{eqnarray}
 &&\left\{C_1(0)\cdots C_8(0) \right\} =
 \left\{\frac{3}{2}, \frac{1}{4},0,-\frac{1}{8},0,\frac{1}{4},0,-\frac{17}{16}\right\}, \nonumber\\
  &&\quad\quad C_n(0)=\frac{E_{n-1}(1)}{2} \quad \text{for $n>1$}.
  \label{eq:cumulants_Kondo}
\end{eqnarray}
with $E_{n}(x)$ being the Euler polynomial.

\subsection{Results and Discussion}

We now analyze different cases in details. We begin with the case of zero degeneracy splitting $\delta=0$.
The dependence of the cumulants on $\kappa$ for $\delta=0$ and $eV\rightarrow 0$ is numerically shown in Fig. \ref{fig:SBMF_Delta0eV0_CUM}.
A recent study based on
SBMF approach~\cite{ChengPRX2014} shows that there is a crossover between Kondo- and Majorana-dominated regimes
as a function of the MZM coupling $\kappa$. For $\kappa\ll \kappa_{c} \equiv \sqrt{T_K/\Gamma} |\epsilon_d|$, the mean field solution is determined by the Kondo temperature $T_K$:
\begin{align}
\tilde {\Gamma}\equiv\Gamma b^2=T_K\equiv\Lambda \exp(-\pi|\epsilon_d|/2\Gamma)
\end{align}
and the renormalized energy level is $\tilde{\epsilon}_d\equiv|\epsilon_d+\eta|\sim \Gamma b^4$.
Here $\Lambda$ is the bandwidth and $\Gamma=\Gamma_R+\Gamma_L$ . Since $b\ll 1$, the renormalized energy $\widetilde{\epsilon}_{d}$ is small $ \widetilde{\epsilon}_{d} \ll \widetilde{\Gamma}$ \cite{ChengPRX2014}. Thus, the cumulants are given by Eq.\eqref{eq:cumulants_Kondo}. In the case of intermediate MZM coupling $\kappa \gg \kappa_c$, the parameter $b\sim \kappa/|\epsilon_d|$ is determined by the Majorana coupling rather than the Kondo temperature. Still, however, if $|\epsilon_d|\gg{\Gamma, \kappa}$, the occupation of the empty state is small $b\ll 1$, and the position of the renormalized level is still close to the Fermi energy $\widetilde{\epsilon}_{d}\sim \Gamma b^4\ll \widetilde{\Gamma}$ \cite{ChengPRX2014}. Thus, the cumulants are given by Eq.\eqref{eq:cumulants_Kondo} for $\Gamma_R=\Gamma_L$.

Next we consider the effect of a finite energy splitting $\delta\neq 0$ and voltage bias on our prediction,
which is important for the experimental detection in realistic settings. The cumulants  $C_n(0)$ for $n=1,2,3,4$ as a function
$\kappa$ and $\delta$ are shown in Fig. \ref{fig:SBMF_CUM}, where we focus on the limit
$|\epsilon_d|\gg\{\Gamma, \kappa\}$. One can see that in order to resolve the universal quantized values,
one has to adjust the voltage bias in the following range 
${\rm min}\{\kappa^2/\Gamma, \Gamma b^2 \}\gg  eV \gg \delta$ where $b=\sqrt{T_K / \Gamma}$ for $\kappa \ll \kappa_c$ 
and $b=\kappa / |\epsilon_d| $ for $\kappa_c \ll \kappa \ll |\epsilon_d|$. The plot of the cumulant power spectra as a 
function of the splitting $\delta$ and Majorana coupling $\kappa$ is shown in Fig. \ref{fig:SBMF_CUM}. 
One can notice that the width of the plateau around the quantized values gradually shrinks with increasing $\delta$.

Finally, we consider strong MZM coupling limit $\kappa\gtrsim |\epsilon_d|$ such that $b\sim 1$ and $\widetilde{\epsilon}_{d}\sim \widetilde{\Gamma}$. In this case, although the energy level shift does not affect the universal values for spin-up channel (due to MZM coupling), the position of the renormalized level $\widetilde{\epsilon}_{d}$ affects the cumulants in the spin-down
channel. Therefore, overall results become nonuniversal and depend on microscopic details such ${\epsilon}_d$ and $\kappa$. As Majorana coupling $\kappa$ is increasing, cumulants start to deviate from the universal values as shown in Fig. \ref{fig:SBMF_Delta0eV0_CUM}.

\section{Conclusions}\label{sec:conclusion}

In this paper, we study the full counting statistics of charge fluctuations in a QD device with a side-coupled TSC as shown in Fig.\ref{fig:setup}. Two normal metal leads, coupled to the QD, are also introduced in order to detect the charge tunneling events. Using a Keldysh path-integral approach, we compute the cumulant generating function for the QD with MZM coupling
in this two lead structure. We first consider a noninteracting spinless system and find that for the symmetric left-right lead couplings, the zero-frequency cumulants exhibit a universal pattern described by a series of numbers generated from the Euler polynomial. This result is independent of the microscopic parameters of the dot, i.e. QD energy level and QD-MZM coupling. For small topological degeneracy splitting due to a finite-size TSC, the universal pattern can also be approximately observed provided the voltage bias is in the appropriate range. We also compute FCS for a spinful QD setup, and consider effect of Coulomb interactions both in the perturbative (weak interactions) and nonperturbative (strong interactions) regimes. In the former case, we find that the interaction-induced corrections to the cumulants appear in the cubic order in voltage $V$ and quadratic order in interactions $U$ indicating that the universal pattern characteristic to Majorana zero modes can be, in principle, measured experimentally. We find that the optimal regime for that corresponds to the strongly-interacting spinful QD in the single-occupancy regime with the intermediate MZM coupling. Our results provide a complete tool set for detecting Majorana modes in tunneling transport measurements which goes far beyond the zero-bias tunneling conductance paradigm.

\subsection*{Acknowledgment}

This work at MSU (A.L.) was supported by NSF Grant No. ECCS-1407875. RL wishes to acknowledge the hospitality of the Aspen Center for Physics and support under NSF Grant \#1066293.

\appendix
\newpage
\begin{widetext}
\section{Derivation of the coefficients in CGF Eq. (\ref{eq:CGF_MZM_finiteT})}\label{app:CGF_coefficients}

In this Appendix, we provide details of the calculations of FCS for non-interacting case.
To evaluate the generating function, we first define
\begin{equation}
 \mathbb{K}(\lambda_L,\lambda_R)=\det\Big[ \mathbb{I}_{4\times 4}-\breve{Q}_{0,dd} \sum_{\alpha} \big( \sum_{i} M_{T,\alpha}^i\otimes \gamma^i \big)^{\dagger}
               \breve{Q}_{0,\alpha} \big( \sum_{i} M_{T,\alpha}^i\otimes \gamma^i \big) \Big].
\end{equation}
We consider a symmetric source-drain bias ($\mu_L=eV/2$ and $\mu_R=-eV/2$), and
insert Eq. (\ref{eq:Q0leads}),  (\ref{eq:Q0dd}), and (\ref{eq:TunningM11}), (\ref{eq:TunningM}) into $ \mathbb{K}(\lambda_L,\lambda_R)$, and obtain
\begin{eqnarray}
  \mathbb{K}(\lambda_L=0,\lambda_R=0) &=& 1 + (\Gamma_L + \Gamma_R)^2 \Big( [G_{0,\bar{d}d}^R]^2 +
    (\Gamma_L + \Gamma_R)^2 [F_{0,dd}^R]^4-
    2 [F_{0,dd}^R]^2 \big(-1 + (\Gamma_L + \Gamma_R)^2 G_{0,d\bar{d}}^R G_{0,\bar{d}d}^R \big) \nonumber\\
    &&+ [G_{0,d\bar{d}}^R]^2 \big(1 + (\Gamma_L + \Gamma_R)^2[ G_{0,\bar{d}d}^R]^2 \big)\Big)
\end{eqnarray}
and
\begin{eqnarray}
  \mathbb{K}(\lambda_L,\lambda_R) &=& \mathbb{K}(0,0) -\mathbb{C}_1 n_L (1-n_L) -\mathbb{C}_2 n_R (1-n_R)
              +\mathbb{B}_1 n_L (1-n_R) +\mathbb{B}_2 n_R (1-n_L) \nonumber\\
              &&+ \mathbb{F} n_L n_R (1-n_L) (1-nR)+ \mathbb{J} n_L n_R (n_L-n_R),
\end{eqnarray}
where
\begin{eqnarray}
 \mathbb{C}_1 &=& 4 e^{-i(\lambda_L+\lambda_R)}\Gamma_L \Gamma_R \Big( 4\Gamma_L \Gamma_R e^{i(\lambda_L-\lambda_R)}
   (e^{i\lambda_L}-e^{i\lambda_R})^2 \big([F_{0,dd}^R]^2 -G_{0,d\bar{d}}^R G_{0,\bar{d}d}^R \big)^2 -(1-e^{i(\lambda_L+\lambda_R)})^2 [F_{0,dd}^R]^2    \Big),\\
  \mathbb{C}_2 &=& 4 e^{-i(\lambda_L+\lambda_R)}\Gamma_L \Gamma_R \Big( 4\Gamma_L \Gamma_R e^{-i(\lambda_L-\lambda_R)}
   (e^{i\lambda_L}-e^{i\lambda_R})^2 \big([F_{0,dd}^R]^2 -G_{0,d\bar{d}}^R G_{0,\bar{d}d}^R \big)^2 -(1-e^{i(\lambda_L+\lambda_R)})^2 [F_{0,dd}^R]^2    \Big),\\
  \mathbb{B}_1 &=& 4 e^{-2 i \lambda _L} \left(-2 e^{i \lambda _L} \left(e^{i \lambda _L}-e^{i \lambda _R}\right) \left([F_{0,dd}^R]^2-G_{0,d\bar{d}}^R G_{0,\bar{d}d}^R\right)^2
\Gamma _L^3 \Gamma _R+e^{2 i \lambda _L} \Big(-1+e^{2 i \lambda _R}\right) [F_{0,dd}^R]^2 \Gamma _R^2 \nonumber\\
       &&  -e^{i \lambda _L} \left(e^{i \lambda _L}-e^{i \lambda _R}\right) \Gamma _L \Gamma _R
       \left([G_{0,d\bar{d}}^R]^2+[G_{0,\bar{d}d}^R]^2+2 \left([F_{0,dd}^R]^2-G_{0,d\bar{d}}^R G_{0,\bar{d}d}^R\right)^2 \Gamma _R^2\right) \nonumber\\
       && +\Gamma_L^2
        \left(-\left(-1+e^{2 i \lambda _L}\right) [F_{0,dd}^R]^2+4 e^{i \lambda _R} \left(-e^{i \lambda _L}
        +e^{i \lambda _R}\right) \left([F_{0,dd}^R]^2-G_{0,d\bar{d}}^R G_{0,\bar{d}d}^R\right)^2 \Gamma _R^2\right)\Big), \\
 \mathbb{B}_2 &=& 4 e^{-2 i \lambda _L} \Big(-2 e^{i \lambda _L} \left(e^{i \lambda _L}-
           e^{i \lambda _R}\right) \left([F_{0,dd}^R]^2-G_{0,d\bar{d}}^R G_{0,\bar{d}d}^R\right)^2 \Gamma _L^3 \Gamma _R \nonumber\\
          &&+e^{2 i \lambda _L} \left(-1+e^{2 i \lambda _R}\right)
            [F_{0,dd}^R]^2 \Gamma _R^2-e^{i \lambda _L} \left(e^{i \lambda _L}-e^{i
             \lambda _R}\right) \Gamma _L \Gamma _R \left([G_{0,d\bar{d}}^R]^2+[G_{0,\bar{d}d}^R]^2+
             2 \left([F_{0,dd}^R]^2 G_{0,d\bar{d}}^R G_{0,\bar{d}d}^R\right)^2 \Gamma _R^2\right)\nonumber\\
          && +\Gamma_L^2 \left(-\left(-1+e^{2 i \lambda _L}\right) [F_{0,dd}^R]^2+
              4 e^{i \lambda _R} \left(-e^{i \lambda _L}+e^{i \lambda _R}\right)
              \left([F_{0,dd}^R]^2-G_{0,d\bar{d}}^R G_{0,\bar{d}d}^R\right)^2 \Gamma _R^2\right)\Big),\\
 \mathbb{F} &=&   256 \Gamma_L^2 \Gamma_R^2 \Big([F_{0,dd}^R]^2- G_{0,d\bar{d}}^R G_{0,\bar{d}d}^R \Big)^2
                      \Big(\sin\frac{\lambda_L-\lambda_R}{2} \Big)^4,\\
 \mathbb{J} &=&   128 i \Gamma_L^2 \Gamma_R^2 \Big([F_{0,dd}^R]^2- G_{0,d\bar{d}}^R G_{0,\bar{d}d}^R \Big)^2
                      \Big(\sin\frac{\lambda_L-\lambda_R}{2} \Big)^2 \sin(\lambda_L-\lambda_R).
\end{eqnarray}
where the Green functions, i.e. $G_{0,\bar{d}d}$,$G_{0,d\bar{d}}$, and $F_{0,dd}$, are defined in Eq. (\ref{eq:G0dd1}) (\ref{eq:G0dd2}) (\ref{eq:F0dd}) of the main text.

\section{Calculation of the CGF corrections due to weak interaction effects}

\subsection{Derivation of the interaction correction formula}
\label{app:Correction}

For the convenience of the calculation, we consider
the polarization function ( see Fig. \ref{fig:DiagramDef} ) in the Schwinger - Keldysh space
(without Larkin-Ovchinnikov rotation)
\begin{eqnarray}
 &&\hat{\Pi}_{\rm P}(\Omega) =
    \begin{pmatrix}
     \hat{\Pi}_{\rm P}^{T}(\Omega) & \hat{\Pi}_{\rm P}^{<}(\Omega) \\
     \hat{\Pi}_{\rm P}^{>}(\Omega) & \hat{\Pi}_{\rm P}^{\widetilde{T}}(\Omega)
    \end{pmatrix} \nonumber\\
  &&= i \int_{-\infty}^{\infty} \frac{d\omega_1}{2\pi}
    \begin{pmatrix}
     G_{d\bar{d}}^{T}(\omega_1+\Omega)G_{d\bar{d}}^{T}(\omega_1) + F_{dd}^{T}(\omega_1+\Omega)F_{\bar{d}\bar{d}}^{T}(\omega_1)\quad\quad
    & G_{d\bar{d}}^{<}(\omega_1+\Omega)G_{d\bar{d}}^{>}(\omega_1) + F_{dd}^{<}(\omega_1+\Omega)F_{\bar{d}\bar{d}}^{>}(\omega_1)\\
    G_{d\bar{d}}^{>}(\omega_1+\Omega)G_{d\bar{d}}^{<}(\omega_1) + F_{dd}^{>}(\omega_1+\Omega)F_{\bar{d}\bar{d}}^{<}(\omega_1)\quad\quad
    & G_{d\bar{d}}^{\widetilde{T}}(\omega_1+\Omega)G_{d\bar{d}}^{\widetilde{T}}(\omega_1) + F_{dd}^{\widetilde{T}}(\omega_1+\Omega)F_{\bar{d}\bar{d}}^{\widetilde{T}}(\omega_1)
    \end{pmatrix}.
\end{eqnarray}
Here the subscript $P$ indicates the particle channel, its particle-hole conjugation (i.e. the hole channel) $\hat{\Pi}_{\rm H}(\Omega)$ has the same form but with replacement $G_{d\bar{d}}\rightarrow G_{\bar{d}d}$ and $F_{dd}\rightarrow F_{\bar{d}\bar{d}}$. The respective self-energy for the spin-up channel can be extracted from
\begin{eqnarray}
 \Sigma_{d\bar{d},\uparrow}^{\lambda_L}(\omega)&=& i U^2 \int \frac{d\Omega}{2\pi}
   \begin{pmatrix}
    G_{d\bar{d},\uparrow}^{T}(\omega-\Omega)\hat{\Pi}^{T}_{\rm P,\downarrow}(\Omega)\quad\quad & G_{d\bar{d},\uparrow}^{<}(\omega-\Omega)\hat{\Pi}^{>}_{\rm P,\downarrow}(\Omega)\\
    G_{d\bar{d},\uparrow}^{>}(\omega-\Omega)\hat{\Pi}^{<}_{\rm P,\downarrow}(\Omega)\quad\quad & G_{d\bar{d},\uparrow}^{\widetilde{T}}(\omega-\Omega)\hat{\Pi}^{\widetilde{T}}_{\rm P,\downarrow}(\Omega)
   \end{pmatrix}\\
 \Sigma_{\bar{d}\bar{d},\uparrow}^{\lambda_L}(\omega)&=& i U^2 \int \frac{d\Omega}{2\pi}
   \begin{pmatrix}
    F_{\bar{d}\bar{d},\uparrow}^{T}(\omega-\Omega)\hat{\Pi}^{T}_{\rm P,\downarrow}(\Omega)\quad\quad & F_{\bar{d}\bar{d},\uparrow}^{<}(\omega-\Omega)\hat{\Pi}^{>}_{\rm P,\downarrow}(\Omega)\\
    F_{\bar{d}\bar{d},\uparrow}^{>}(\omega-\Omega)\hat{\Pi}^{<}_{\rm P,\downarrow}(\Omega)\quad\quad & F_{\bar{d}\bar{d},\uparrow}^{\widetilde{T}}(\omega-\Omega)\hat{\Pi,\downarrow}^{\widetilde{T}}_{\rm P}(\Omega)
   \end{pmatrix}\\
\Sigma_{dd,\uparrow}^{\lambda_L}(\omega)&=& i U^2 \int \frac{d\Omega}{2\pi}
   \begin{pmatrix}
    F_{dd,\uparrow}^{T}(\omega-\Omega)\hat{\Pi}^{T}_{\rm H,\downarrow}(\Omega)\quad\quad & F_{dd,\uparrow}^{<}(\omega-\Omega)\hat{\Pi}^{>}_{\rm H,\downarrow}(\Omega)\\
    F_{dd,\uparrow}^{>}(\omega-\Omega)\hat{\Pi}^{<}_{\rm H,\downarrow}(\Omega)\quad\quad & F_{dd,\uparrow}^{\widetilde{T}}(\omega-\Omega)\hat{\Pi}^{\widetilde{T}}_{\rm H,\downarrow}(\Omega)
   \end{pmatrix}\\
\Sigma_{\bar{d}d,\uparrow}^{\lambda_L}(\omega)&=& i U^2 \int \frac{d\Omega}{2\pi}
   \begin{pmatrix}
    G_{\bar{d}d,\uparrow}^{T}(\omega-\Omega)\hat{\Pi}^{T}_{\rm H,\downarrow}(\Omega)\quad\quad & G_{\bar{d}d,\uparrow}^{<}(\omega-\Omega)\hat{\Pi}^{>}_{\rm H,\downarrow}(\Omega)\\
    G_{\bar{d}d,\uparrow}^{>}(\omega-\Omega)\hat{\Pi}^{<}_{\rm H,\downarrow}(\Omega)\quad\quad & G_{\bar{d}d,\uparrow}^{\widetilde{T}}(\omega-\Omega)\hat{\Pi}^{\widetilde{T}}_{\rm H,\downarrow}(\Omega)
   \end{pmatrix}
\end{eqnarray}
where the spin-up channel couples to MZM and the spin-down channel does not. We want to calculate the following function, i.e. interaction correction in Eq.(\ref{eq:C_interaction})
\begin{eqnarray}
 &&\int_{-\infty}^{\infty} \frac{d \omega}{2\pi} \rm{Tr}\big( \breve{Q}_{dd,\uparrow,U=0}^{\lambda_L}(\omega) \Sigma_{\uparrow}^{\lambda_L}(\omega) \big)\nonumber\\
   && = \int_{-\infty}^{\infty} \frac{d \omega}{2\pi} \rm{Tr}\Big( G_{d\bar{d},\uparrow}(\omega) \Sigma_{d\bar{d},\uparrow}^{\lambda_L}(\omega)
          + F_{dd,\uparrow}(\omega) \Sigma_{\bar{d}\bar{d},\uparrow}^{\lambda_L}(\omega)
           + F_{\bar{d}\bar{d},\uparrow}(\omega) \Sigma_{dd,\uparrow}^{\lambda_L}(\omega)+ G_{\bar{d}d,\uparrow}(\omega) \Sigma_{\bar{d}d,\uparrow}^{\lambda_L}(\omega)\Big)\nonumber\\
   && = \int_{-\infty}^{\infty} \frac{d \omega}{2\pi} \rm{Tr}
               \Big(\gamma^{cl} G_{d\bar{d},\uparrow}(\omega) \gamma^{cl} \Sigma_{d\bar{d},\uparrow}^{\lambda_L}(\omega)
               + \gamma^{cl} F_{dd,\uparrow}(\omega) \gamma^{cl} \Sigma_{\bar{d}\bar{d},\uparrow}^{\lambda_L}(\omega)
           + \gamma^{cl} F_{\bar{d}\bar{d},\uparrow}(\omega) \gamma^{cl} \Sigma_{dd,\uparrow}^{\lambda_L}(\omega)+
           \gamma^{cl} G_{\bar{d}d,\uparrow}(\omega) \gamma^{cl}\Sigma_{\bar{d}d,\uparrow}^{\lambda_L}(\omega)\Big)\nonumber\\
\end{eqnarray}
where $\gamma^{cl}=\mathbb{I}_{2\times 2}$. Note that the matrix Green functions and self-energies (which are from Eq.(\ref{eq:C_interaction})) have the form
$\bigl(\begin{smallmatrix}
G^{R}&G^{K}\\ G^{\widetilde{K}}&G^{A}
\end{smallmatrix} \bigr)$
(i.e. with the L-O rotation). Therefore, the function above just corresponds to the classical-classical part ($\widetilde{K}$)
of a certain polarization function, and we define a function $\Xi_{\lambda,\sigma}^{\widetilde{K}}$ for interaction correction
\begin{equation}
 U^2\; \Xi_{\lambda,\uparrow}^{\widetilde{K}} =
  \int_{-\infty}^{\infty} \frac{d \omega}{2\pi} \rm{Tr}
               \Big(\gamma^{cl} G_{d\bar{d},\uparrow}(\omega) \gamma^{cl} \Sigma_{d\bar{d},\uparrow}^{\lambda_L}(\omega)
               + \gamma^{cl} F_{dd,\uparrow}(\omega) \gamma^{cl} \Sigma_{\bar{d}\bar{d},\uparrow}^{\lambda_L}(\omega)
           + \gamma^{cl} F_{\bar{d}\bar{d},\uparrow}(\omega) \gamma^{cl} \Sigma_{dd,\uparrow}^{\lambda_L}(\omega)+
           \gamma^{cl} G_{\bar{d}d,\uparrow}(\omega) \gamma^{cl}\Sigma_{\bar{d}d,\uparrow}^{\lambda_L}(\omega)\Big),
\end{equation}
which can be described by the diagrams shown in Fig. \ref{fig:CGFUDiag22}.
The function $\Xi_{\lambda}$ in the Schwinger - Keldysh space (without L-O rotation) has the simple form
\begin{equation}
\Xi_{\lambda} =
 \begin{pmatrix}
  \Xi_{\lambda}^{T} & \Xi_{\lambda}^{<} \\
  \Xi_{\lambda}^{>} & \Xi_{\lambda}^{\widetilde{T}}
 \end{pmatrix},
\end{equation}
where
\begin{eqnarray}
\Xi_{\lambda}^{T} &=& \int_{-\infty}^{\infty} \frac{d\Omega}{2\pi} \Big( \hat{\Pi}_{P,\uparrow}^{T}(-\Omega)\hat{\Pi}_{P,\downarrow}^{T}(\Omega)
                        +\hat{\Pi}_{H,\uparrow}^{T}(-\Omega)\hat{\Pi}_{H,\downarrow}^{T}(\Omega) \Big),\\
\Xi_{\lambda}^{\widetilde{T}} &=& \int_{-\infty}^{\infty} \frac{d\Omega}{2\pi} \Big( \hat{\Pi}_{P,\uparrow}^{\widetilde{T}}(-\Omega)\hat{\Pi}_{P,\downarrow}^{\widetilde{T}}(\Omega)
                        +\hat{\Pi}_{H,\uparrow}^{\widetilde{T}}(-\Omega)\hat{\Pi}_{H,\downarrow}^{\widetilde{T}}(\Omega) \Big),\\
\Xi_{\lambda}^{<} &=& \int_{-\infty}^{\infty} \frac{d\Omega}{2\pi} \Big( \hat{\Pi}_{P,\uparrow}^{<}(-\Omega)\hat{\Pi}_{P,\downarrow}^{<}(\Omega)
                        +\hat{\Pi}_{H,\uparrow}^{<}(-\Omega)\hat{\Pi}_{H,\downarrow}^{<}(\Omega) \Big),\\
\Xi_{\lambda}^{>} &=& \int_{-\infty}^{\infty} \frac{d\Omega}{2\pi} \Big( \hat{\Pi}_{P,\uparrow}^{>}(-\Omega)\hat{\Pi}_{P,\downarrow}^{>}(\Omega)
                        +\hat{\Pi}_{H,\uparrow}^{>}(-\Omega)\hat{\Pi}_{H,\downarrow}^{>}(\Omega) \Big),
\end{eqnarray}
and
\begin{equation}
 \Xi_{\lambda}^{\widetilde{K}}=\Big(\Xi_{\lambda}^{T}+\Xi_{\lambda}^{\widetilde{T}}-\Xi_{\lambda}^{<}-\Xi_{\lambda}^{>}\Big)/2.
\end{equation}
Finally, we reach the formula in Eq.(\ref{eq:C_interaction_Formula}) of the main text.

\end{widetext}

\subsection{Expansion of $\Xi_{\lambda}^{\widetilde{K}}$}\label{app:Expansion}

Now, let's look at how the correction $\Xi_{\lambda}^{\widetilde{K}}$ changes as a function of source-drain bias $eV$ for $T=0$.
For small $eV$, we can expand the function
\begin{eqnarray}
 \Xi_{\lambda}^{\widetilde{K}}(eV)&=&\Xi_{\lambda}^{\widetilde{K}}(0)+(eV) \Xi_{\lambda}^{\widetilde{K},(1)}(0)
       +(eV)^2 \Xi_{\lambda}^{\widetilde{K},(2)}(0)\nonumber\\
       &&+(eV)^3 \Xi_{\lambda}^{\widetilde{K},(3)}(0)+\cdots
\end{eqnarray}
Due to the causality reasons (this is still true in the presence of artificial counting field),
the polarization function at $T=0$ and $eV=0$ has the following properties: $\hat{\Pi}_{P,\sigma}^{<}(\Omega)\propto n(\Omega)$ and
 $\hat{\Pi}_{P,\sigma}^{>}(\Omega)\propto 1-n(\Omega)$, where $n(\Omega)=1-\theta(\Omega)$ is the Fermi-distribution function
at $T=0$. Then, we find $\Xi_{\lambda}^{<}(eV=0)=0$ and $\Xi_{\lambda}^{>}(eV=0)=0$ for $T=0$.
Although the time-ordered and anti-time-ordered parts of the Green functions
($G_{d\bar{d}}^{T,\widetilde{T}}$,$G_{\bar{d}d}^{T,\widetilde{T}}$, $F_{dd}^{T,\widetilde{T}}$, and $F_{\bar{d}\bar{d}}^{T,\widetilde{T}}$)
depends on the counting field $\lambda_L$, their dependence on $\lambda_L$ enters in a way such that
all the $\lambda_L$ dependent terms have a pre-factor $n_{L} (1-n_{R})$. At $T=0$ and $eV=0$,  $n_{L} (1-n_{R})$ vanishes
and time-order and anti-time-order Green functions are independent of $\lambda_L$. Therefore, we have
$\Xi_{\lambda}^{T}(eV=0)=\Xi_{\lambda=0}^{T}(eV=0)$ and $\Xi_{\lambda}^{\widetilde{T}}(eV=0)=\Xi_{\lambda=0}^{\widetilde{T}}(eV=0)$.
Combining those relations with $\Xi_{\lambda=0}^{\widetilde{K}}(eV=0)=\Xi_{\lambda=0}^{T}(eV=0)+\Xi_{\lambda=0}^{\widetilde{T}}(eV=0)=0$,
we can show
\begin{equation}
 \Xi_{\lambda}^{\widetilde{K}}(0) = 0.
\end{equation}
This is intuitively obvious since there should be no current in equilibrium (i.e. $eV\rightarrow 0$).

To consider higher order terms, we expand the integrand of the polarization function integral in order of $eV$.
These integrands have the following form
\begin{eqnarray}
 &F^{\alpha}(\omega_1,\Omega | n_L(\omega_1+\Omega),n_L(\omega_1),n_R(\omega_1+\Omega),n_R(\omega_1)) \nonumber\\
 &=G_{d\bar{d}}^{\alpha}(\omega_1+\Omega)G_{d\bar{d}}^{\alpha}(\omega_1) + F_{dd}^{\alpha}(\omega_1+\Omega)F_{\bar{d}\bar{d}}^{\alpha}(\omega_1),
 \label{eq:F_def}
\end{eqnarray}
where $\alpha=T,\widetilde{T},<,>$,
The bias $eV$ only enters through the Fermi distribution function $n_L$ and $n_R$: $n_L(\omega)=1-\theta(\omega-eV/2)$,
$n_R(\omega)=1-\theta(\omega+eV/2)$.
Due to the properties of Heaviside theta function, by expansion and resummation, one can prove the following relation:
If one has a series of functions $n_i(\omega)=1-\theta(\omega-\omega_i)$
with $i=1,2,\cdots,k$ and $\omega_1\leqslant\omega_2\leqslant\cdots \leqslant\omega_k$, then
\begin{eqnarray}
 &&F(n_1,n_2,\cdots,n_k)=  \nonumber\\
  &&F(0,0,\cdots,0)+ \Big[F(1,1,\cdots,1)-F(0,1,\cdots,1) \Big] n_1 +\cdots \nonumber\\
          && + \Big[F(\overbrace{0,\cdots,0}^{i},1,\cdots,1)-F(\overbrace{0,\cdots,0}^{i+1},1,\cdots,1) \Big] n_i +\cdots\nonumber\\
          &&   + \Big[F(0,\cdots,0,1)-F(0,0,\cdots,0) \Big] n_k.
          \label{eq:FT_linearizeN}
\end{eqnarray}

Note that this formula depends on the order of the arguments in the Fermi distribution function. We then define the polarization function for different regions where the function $F$ has different forms. First of all, we consider $\Omega\geq eV$ such that $-eV/2-\Omega< eV/2-\Omega\leq -eV/2<eV/2$, and define
\begin{equation}
 \hat{\Pi}_{P1,\sigma}^{\alpha}(\Omega,eV)=i\int_{-\infty}^{\infty} \frac{d\omega_1}{2\pi} F_{1}^{\alpha}(\omega_1,\Omega),
\end{equation}
and
$F_{1}^{\alpha}(\omega_1,\Omega|n_{R}(\omega_1+\Omega),n_{L}(\omega_1+\Omega),n_{R}(\omega_1),n_{L}(\omega_1))=
F^{\alpha}(\omega_1,\Omega|n_{L}(\omega_1+\Omega),n_{L}(\omega_1),n_{R}(\omega_1+\Omega),n_{R}(\omega_1))$,
where $F^{\alpha}$ is the same function as the one defined in Eq. (\ref{eq:F_def}),
but the arguments in the function $F_1^{\alpha}$ have the different order.
Secondly, we consider $0<\Omega< eV$ such that
$-eV/2-\Omega< -eV/2< eV/2-\Omega <eV/2$, and define
\begin{equation}
 \hat{\Pi}_{P2,\sigma}^{\alpha}(\Omega,eV)=i\int_{-\infty}^{\infty} \frac{d\omega_1}{2\pi} F_{2}^{\alpha}(\omega_1,\Omega),
 \label{eq:Pi2_Def}
\end{equation}
where
$F_{2}^{\alpha}(\omega_1,\Omega|n_{R}(\omega_1+\Omega),n_{R}(\omega_1),n_{L}(\omega_1+\Omega),n_{L}(\omega_1))=
F^{\alpha}(\omega_1,\Omega|n_{L}(\omega_1+\Omega),n_{L}(\omega_1),n_{R}(\omega_1+\Omega),n_{R}(\omega_1))$.
Thirdly, we consider $-eV<\Omega< 0$ such that
$-eV/2< -eV/2-\Omega< eV/2 <eV/2-\Omega$, and define
\begin{equation}
 \hat{\Pi}_{P3,\sigma}^{\alpha}(\Omega,eV)=i\int_{-\infty}^{\infty} \frac{d\omega_1}{2\pi} F_{3}^{\alpha}(\omega_1,\Omega),
\end{equation}
where $F_{3}^{\alpha}(\omega_1,\Omega|n_{R}(\omega_1),n_{R}(\omega_1+\Omega),n_{L}(\omega_1),n_{L}(\omega_1+\Omega))
= F^{\alpha}(\omega_1,\Omega|n_{L}(\omega_1+\Omega),n_{L}(\omega_1),n_{R}(\omega_1+\Omega),n_{R}(\omega_1))$.
Finally, we consider $\Omega<-eV$ such that
$-eV/2< eV/2< -eV/2-\Omega <eV/2-\Omega$, and define
\begin{equation}
 \hat{\Pi}_{P4,\sigma}^{\alpha}(\Omega,eV)=i\int_{-\infty}^{\infty} \frac{d\omega_1}{2\pi} F_{4}^{\alpha}(\omega_1,\Omega),
\end{equation}
where $F_{4}^{\alpha}(\omega_1,\Omega|n_{R}(\omega_1),n_{L}(\omega_1),n_{R}(\omega_1+\Omega),n_{L}(\omega_1+\Omega)) =
 F^{\alpha}(\omega_1,\Omega|n_{L}(\omega_1+\Omega),n_{L}(\omega_1),n_{R}(\omega_1+\Omega),n_{R}(\omega_1))$.

\begin{widetext}
Following the definition above, the correction $\Xi_{\lambda}^{\widetilde{K}}$ can be written as (here we consider $\epsilon_d=0$)
\begin{equation}
 \Xi_{\lambda}^{\alpha} = \Xi_{\lambda,A}^{\alpha}+\Xi_{\lambda,B}^{\alpha},
\end{equation}
where
\begin{eqnarray}
 \Xi_{\lambda,A}^{\alpha} &=& 2\int_{0}^{\infty} \frac{d\Omega}{2\pi} \hat{\Pi}_{P1,\uparrow}^{\alpha}(\Omega) \hat{\Pi}_{P4,\downarrow}(-\Omega)^{\alpha}+
                       2\int_{-\infty}^{0} \frac{d\Omega}{2\pi} \hat{\Pi}_{P4,\uparrow}^{\alpha}(\Omega)
                            \hat{\Pi}_{P1,\downarrow}^{\alpha}(-\Omega)\nonumber\\
 \Xi_{\lambda,B}^{\alpha} &=&2\int_{0}^{eV} \frac{d\Omega}{2\pi} \Big[\hat{\Pi}_{P2,\uparrow}^{\alpha}(\Omega) \hat{\Pi}_{P3,\downarrow}(-\Omega)^{\alpha}
                                -\hat{\Pi}_{P1,\uparrow}^{\alpha}(\Omega) \hat{\Pi}_{P4,\downarrow}(-\Omega)^{\alpha} \Big] \nonumber\\
                     &&  + 2\int_{-eV}^{0} \frac{d\Omega}{2\pi} \Big[\hat{\Pi}_{P3,\uparrow}^{\alpha}(\Omega) \hat{\Pi}_{P2,\downarrow}(-\Omega)^{\alpha}
                                -\hat{\Pi}_{P4,\uparrow}^{\alpha}(\Omega) \hat{\Pi}_{P1,\downarrow}(-\Omega)^{\alpha} \Big] \nonumber\\
                     &=& \Xi_{\lambda,A}^{\alpha} + \Xi_{\lambda,B}^{\alpha}.
                     \label{eq:Correction_Formula}
\end{eqnarray}
Here the factor 2 comes from the summation of both particle and hole part
(note that $\hat{\Pi}_{P,\sigma}=\hat{\Pi}_{H,\sigma}$ for $\epsilon_d=0$).
The whole correction includes two parts: the first part $\Xi_{\lambda,A}^{\alpha}$ and a leftover part $\Xi_{\lambda,B}^{\alpha}$.
To check this formalism, we considered the case without MZM coupling and reproduce the result shown in Eq. (21) of
Gogolin and Komnik \cite{Gogolin&Komnik06}. Note, in order to obtain the right result, we need to take appropriate order of limits.
If we want to reproduce the $\kappa=0$ result, we have to take the limit $\kappa\rightarrow 0$ before taking the
$\omega\rightarrow 0$ limit for linear response. Then, we focus on the case with MZM coupling (where
if we want to consider zero Majorana splitting, i.e. $\delta=0$, we have to take the limit $\delta\rightarrow 0$ first
before taking the $\omega\rightarrow 0$ limit for linear response).

\subsection{Derivation of the first part of interaction correction $\Xi_{\lambda,A}^{\widetilde{K},(2)}$}
\label{app:Derivation_XiA}

First of all, we focus on the first part $\Xi_{\lambda,A}^{\alpha}$, and  expand it
in order of $eV$.
We also notice that after the transformation of Eq. (\ref{eq:FT_linearizeN}), the integrands of the polarization
functions are linear in the Fermi distribution function. Therefore, the expansion can be obtained analytically
by expanding the Fermi distribution function
\begin{equation}
 n(\omega\pm\frac{eV}{2})=n(\omega) -\delta(\omega) \Big( \pm\frac{eV}{2} \Big)-\frac{1}{2!}\delta^{'}(\omega) \Big( \pm\frac{eV}{2} \Big)^2
           -\frac{1}{3!}\delta^{''}(\omega) \Big( \pm\frac{eV}{2} \Big)^3+\cdots
\end{equation}

After expanding the Fermi distribution function in order of $eV$, we further expand the  polarization function
\begin{equation}
 \hat{\Pi}_{P}(\Omega,eV)= \hat{\Pi}_{P}(\Omega,0) + (eV) \hat{\Pi}_{P}^{(1)}(\Omega,0)+ (eV)^2 \hat{\Pi}_{P}^{(2)}(\Omega,0)
                              + (eV)^3 \hat{\Pi}_{P}^{(3)}(\Omega,0)+\cdots
 \end{equation}
After the integration by part for the Dirac-delta function, the linear terms can be obtained
\begin{eqnarray}
 \hat{\Pi}_{P}^{\alpha,(1)}(\Omega>0,0) &=& \frac{i}{4\pi} \Big[ - F_{1}^{\alpha}(-\Omega,\Omega|1,1,1,1)
                                             +2 F_{1}^{\alpha}(-\Omega,\Omega|0,1,1,1)-F_{1}^{\alpha}(-\Omega,\Omega|0,0,1,1) \nonumber\\
                             &&  - F_{1}^{\alpha}(0,\Omega|0,0,1,1)
                                             +2 F_{1}^{\alpha}(0,\Omega|0,0,0,1)-F_{1}^{\alpha}(0,\Omega|0,0,0,0) \Big],\\
\hat{\Pi}_{P}^{\alpha,(1)}(\Omega<0,0) &=& \frac{i}{4\pi} \Big[ - F_{4}^{\alpha}(0,\Omega|1,1,1,1)
                                             +2 F_{4}^{\alpha}(0,\Omega|0,1,1,1)-F_{4}^{\alpha}(0,\Omega|0,0,1,1) \nonumber\\
                            &&   - F_{4}^{\alpha}(-\Omega,\Omega|0,0,1,1)
                                             +2 F_{4}^{\alpha}(-\Omega,\Omega|0,0,0,1)-F_{4}^{\alpha}(-\Omega,\Omega|0,0,0,0) \Big],
\end{eqnarray}
where $\alpha=T,\widetilde{T},<,>$.
After simplification of those expression above, we find for both spin-channels at $T=0$
\begin{equation}
 \hat{\Pi}_{P}^{T,(1)}(\Omega,0)=0, \quad\quad\quad \hat{\Pi}_{P}^{\widetilde{T},(1)}(\Omega,0)=0, \quad\quad\quad
 \hat{\Pi}_{P}^{<,(1)}(\Omega,0)\sim \theta(-\Omega), \quad\quad\quad \hat{\Pi}_{P}^{>,(1)}(\Omega,0)\sim \theta(\Omega).
\end{equation}
We then obtain the linear correction to the generating function
\begin{eqnarray}
 \Xi_{\lambda}^{\widetilde{K},(1)}(0) &=& 2 \int_{-\infty}^{\infty} \frac{d \Omega}{2\pi} \Big[ \hat{\Pi}_{P,\uparrow}^{T}(-\Omega,0) \hat{\Pi}_{P,\downarrow}^{T,(1)}(\Omega,0)
                           + \hat{\Pi}_{P,\uparrow}^{\widetilde{T}}(-\Omega,0) \hat{\Pi}_{P,\downarrow}^{\widetilde{T},(1)}(\Omega,0)\nonumber\\
         && \quad\quad\quad\quad        + \hat{\Pi}_{P,\uparrow}^{<}(-\Omega,0) \hat{\Pi}_{P,\downarrow}^{<,(1)}(\Omega,0)
                           + \hat{\Pi}_{P,\uparrow}^{>}(-\Omega,0) \hat{\Pi}_{P,\downarrow}^{>,(1)}(\Omega,0)\nonumber\\
       && \quad\quad\quad\quad  +  \hat{\Pi}_{P,\uparrow}^{T,(1)}(-\Omega,0) \hat{\Pi}_{P,\downarrow}^{T}(\Omega,0)
                           + \hat{\Pi}_{P,\uparrow}^{\widetilde{T},(1)}(-\Omega,0) \hat{\Pi}_{P,\downarrow}^{\widetilde{T}}(\Omega,0)\nonumber\\
        && \quad\quad\quad\quad          + \hat{\Pi}_{P,\uparrow}^{<,(1)}(-\Omega,0) \hat{\Pi}_{P,\downarrow}^{<}(\Omega,0)
                           + \hat{\Pi}_{P,\uparrow}^{>,(1)}(-\Omega,0) \hat{\Pi}_{P,\downarrow}^{>}(\Omega,0) \Big]\nonumber\\
       & =& 0.
\end{eqnarray}
Similarly, the quadratic terms can be written as
\begin{eqnarray}
 \hat{\Pi}_{P}^{\alpha,(2)}(\Omega>0,0) &=& \frac{i}{16\pi}\Big[ \partial_{\omega_1}F_{1}^{\alpha}(\omega_1=-\Omega,\Omega|1,1,1,1)
                     -\partial_{\omega_1}F_{1}^{\alpha}(\omega_1=-\Omega,\Omega|0,0,1,1) \nonumber\\
                    &&\quad\quad\quad\quad\quad +\partial_{\omega_1}F_{1}^{\alpha}(\omega_1=0,\Omega|0,0,1,1)
                      -\partial_{\omega_1}F_{1}^{\alpha}(\omega_1=0,\Omega|0,0,0,0)  \Big],\\
 \hat{\Pi}_{P}^{\alpha,(2)}(\Omega<0,0) &=& \frac{i}{16\pi}\Big[ \partial_{\omega_1}F_{4}^{\alpha}(\omega_1=-\Omega,\Omega|1,1,1,1)
                     -\partial_{\omega_1}F_{4}^{\alpha}(\omega_1=-\Omega,\Omega|0,0,1,1) \nonumber\\
                    &&\quad\quad\quad\quad\quad +\partial_{\omega_1}F_{4}^{\alpha}(\omega_1=0,\Omega|0,0,1,1)
                      -\partial_{\omega_1}F_{4}^{\alpha}(\omega_1=0,\Omega|0,0,0,0)  \Big].
\end{eqnarray}
The quadratic term of the generating function is therefore
\begin{eqnarray}
 \Xi_{\lambda}^{\widetilde{K},(2)}(0) &=& 2 \int_{-\infty}^{\infty} \frac{d \Omega}{2\pi}
                         \Big[ \hat{\Pi}_{P,\uparrow}^{T}(-\Omega,0) \hat{\Pi}_{P,\downarrow}^{T,(2)}(\Omega,0)
                  + \hat{\Pi}_{P,\uparrow}^{\widetilde{T}}(-\Omega,0) \hat{\Pi}_{P,\downarrow}^{\widetilde{T},(2)}(\Omega,0)\nonumber\\
       && \quad\quad\quad\quad  +  \hat{\Pi}_{P,\uparrow}^{T,(2)}(-\Omega,0) \hat{\Pi}_{P,\downarrow}^{T}(\Omega,0)
                           + \hat{\Pi}_{P,\uparrow}^{\widetilde{T},(2)}(-\Omega,0) \hat{\Pi}_{P,\downarrow}^{\widetilde{T}}(\Omega,0)\Big],
\end{eqnarray}
where we use the relations $\hat{\Pi}^{>}(-\Omega,0)\hat{\Pi}^{>}(\Omega,0)=0$, $\hat{\Pi}^{<}(-\Omega,0)\hat{\Pi}^{<}(\Omega,0)=0$,
$\hat{\Pi}^{(1),T}(-\Omega,0)=0$, and $\hat{\Pi}^{(1),\widetilde{T}}(-\Omega,0)=0$.
After the simplification of polarization functions , we find that
the time-ordered and anti-time ordered parts $\hat{\Pi}_{P}^{\alpha,(2)}(\Omega,0)$ (for $\alpha=T, \widetilde{T}$) do not
depend on the counting field $\lambda_L$. Then, we conclude that the quadratic term $\Xi_{\lambda}^{\widetilde{K},(2)}(0)$
is the same as the term with $\lambda_L=0$, which is zero due to causality and unitarity
\begin{equation}
  \Xi_{\lambda}^{\widetilde{K},(2)}(0) = 0.
\end{equation}
The cubic term of the polarization function corresponds to the integral of $\delta^{''}$, and thus reads
\begin{eqnarray}
 \hat{\Pi}_{P}^{\alpha,(3)}(\Omega>0,0) &=& \frac{i}{3! 2^3 2\pi}\Big[ -\partial^2_{\omega_1}F_{1}^{\alpha}(-\Omega,\Omega|1,1,1,1)
                    +2\partial^2_{\omega_1}F_{1}^{\alpha}(-\Omega,\Omega|0,1,1,1)  -\partial^2_{\omega_1}F_{1}^{\alpha}(-\Omega,\Omega|0,0,1,1) \nonumber\\
                    &&\quad\quad -\partial^2_{\omega_1}F_{1}^{\alpha}(0,\Omega|0,0,1,1) +2\partial^2_{\omega_1}F_{1}^{\alpha}(0,\Omega|0,0,1,1)
                      -\partial^2_{\omega_1}F_{1}^{\alpha}(0,\Omega|0,0,0,0)  \Big],\\
 \hat{\Pi}_{P}^{\alpha,(3)}(\Omega<0,0) &=& \frac{i}{3! 2^3 2\pi}\Big[ -\partial^2_{\omega_1}F_{4}^{\alpha}(0,\Omega|1,1,1,1)
                    +2\partial^2_{\omega_1}F_{4}^{\alpha}(0,\Omega|0,1,1,1)  -\partial^2_{\omega_1}F_{4}^{\alpha}(0,\Omega|0,0,1,1) \nonumber\\
                    &&\quad\quad -\partial^2_{\omega_1}F_{4}^{\alpha}(-\Omega,\Omega|0,0,1,1) +2\partial^2_{\omega_1}F_{4}^{\alpha}(-\Omega,\Omega|0,0,1,1)
                      -\partial^2_{\omega_1}F_{4}^{\alpha}(-\Omega,\Omega|0,0,0,0)  \Big].
\end{eqnarray}
We evaluate and simplify the function above for $\epsilon_d=0$ and $\Gamma_L=\Gamma_R=\Gamma/2$, and obtain
\begin{eqnarray}
  \hat{\Pi}_{P,\uparrow}^{T,(3)}(\Omega,0) &=& \frac{-i (e^{i\lambda_L}-1) \big[(e^{i\lambda_L}-1)\Gamma^2-2(e^{i\lambda_L}+3)\kappa^2 \big]}
                                                 {48\pi (e^{i\lambda_L}+1)^2\Gamma^2\kappa^2} \frac{1}{(|\Omega|+i\Gamma)^2},\\
  \hat{\Pi}_{P,\uparrow}^{\widetilde{T},(3)}(\Omega,0) &=& \frac{-i (e^{i\lambda_L}-1) \big[(e^{i\lambda_L}-1)\Gamma^2-2(e^{i\lambda_L}+3)\kappa^2 \big]}
                                                 {48\pi (e^{i\lambda_L}+1)^2\Gamma^2\kappa^2} \frac{1}{(|\Omega|-i\Gamma)^2},\\
  \hat{\Pi}_{P,\downarrow}^{T,(3)}(\Omega,0) &=& \frac{-i (e^{-i\lambda_L}-1)}{12\pi \Gamma^2} \frac{1}{(|\Omega|+i\Gamma)^2},\\
  \hat{\Pi}_{P,\downarrow}^{\widetilde{T},(3)}(\Omega,0) &=& \frac{-i (e^{-i\lambda_L}-1)}{12\pi \Gamma^2} \frac{1}{(|\Omega|-i\Gamma)^2}.
\end{eqnarray}
The cubic term of the generating function reads
\begin{eqnarray}
 \Xi_{\lambda,A}^{\widetilde{K},(3)}(0) &=& 2 \int_{-\infty}^{\infty} \frac{d \Omega}{2\pi}
                         \Big[ \hat{\Pi}_{P,\uparrow}^{T}(-\Omega,0) \hat{\Pi}_{P,\downarrow}^{T,(3)}(\Omega,0)
                  + \hat{\Pi}_{P,\uparrow}^{\widetilde{T}}(-\Omega,0) \hat{\Pi}_{P,\downarrow}^{\widetilde{T},(3)}(\Omega,0)\nonumber\\
       && \quad\quad\quad\quad  +  \hat{\Pi}_{P,\uparrow}^{T,(3)}(-\Omega,0) \hat{\Pi}_{P,\downarrow}^{T}(\Omega,0)
                           + \hat{\Pi}_{P,\uparrow}^{\widetilde{T},(3)}(-\Omega,0) \hat{\Pi}_{P,\downarrow}^{\widetilde{T}}(\Omega,0)\Big]\nonumber\\
       &=& \frac{-i (e^{-i\lambda_L}-1)}{6\pi \Gamma^2} \int_{-\infty}^{\infty} \frac{d \Omega}{2\pi} \Big[
             \hat{\Pi}_{P,\uparrow}^{T}(-\Omega,0) \frac{1}{(|\Omega|+i\Gamma)^2} + \hat{\Pi}_{P,\uparrow}^{\widetilde{T}}(-\Omega,0) \frac{1}{(|\Omega|-i\Gamma)^2}\Big]   \nonumber\\
       &&+ \frac{-i (e^{i\lambda_L}-1) \big[(e^{i\lambda_L}-1)\Gamma^2-2(e^{i\lambda_L}+3)\kappa^2 \big]} {24\pi (e^{i\lambda_L}+1)^2\Gamma^2\kappa^2} \nonumber\\
       && \times  \int_{-\infty}^{\infty} \frac{d \Omega}{2\pi} \Big[\hat{\Pi}_{P,\downarrow}^{T}(-\Omega,0) \frac{1}{(|\Omega|+i\Gamma)^2}
           + \hat{\Pi}_{P,\downarrow}^{\widetilde{T}}(-\Omega,0) \frac{1}{(|\Omega|-i\Gamma)^2}\Big] \nonumber\\
        &=& \frac{ (e^{-i\lambda_L}-1)}{6\pi \Gamma^4} \mathbf{O}(\widetilde{\kappa})
       + (2-\frac{\pi^2}{4}) \frac{(e^{i\lambda_L}-1) \big[(e^{i\lambda_L}-1)-2(e^{i\lambda_L}+3)\widetilde{\kappa}^2 \big]}
       {12\pi^3 (e^{i\lambda_L}+1)^2\Gamma^4 \widetilde{\kappa}^2},
\end{eqnarray}
where
\begin{equation}
 \mathbf{O}(\widetilde{\kappa}) = \int_{-\infty}^{\infty} \frac{d \widetilde{\Omega}}{2\pi} \Big[
             \hat{\Pi}_{P,\uparrow}^{T}(-\widetilde{\Omega},0) \frac{-i}{(|\widetilde{\Omega}|+i)^2}
             + \hat{\Pi}_{P,\uparrow}^{\widetilde{T}}(-\widetilde{\Omega},0) \frac{-i}{(|\widetilde{\Omega}|-i)^2}\Big],
              \text{with $\widetilde{\Omega}=\frac{\Omega}{\Gamma}$, $\widetilde{\kappa}=\frac{\kappa}{\Gamma}$.}
\end{equation}
This correction is nonzero for $\lambda_L\neq 0$.

\subsection{Derivation of the second part of interaction correction $\Xi_{\lambda,B}^{\widetilde{K},(2)}$}
\label{app:Derivation_XiB}

Now, let's consider the leftover terms
\begin{eqnarray}
&&2\int_{0}^{eV} \frac{d\Omega}{2\pi} \Big[\hat{\Pi}_{P2,\uparrow}^{\alpha}(\Omega,eV) \hat{\Pi}_{P3,\downarrow}^{\alpha}(-\Omega,eV)
                                -\hat{\Pi}_{P1,\uparrow}^{\alpha}(\Omega,eV) \hat{\Pi}_{P4,\downarrow}^{\alpha}(-\Omega,eV) \Big] \nonumber\\
&& + 2\int_{-eV}^{0} \frac{d\Omega}{2\pi} \Big[\hat{\Pi}_{P3,\uparrow}^{\alpha}(\Omega,eV) \hat{\Pi}_{P2,\downarrow}^{\alpha}(-\Omega,eV)
                                -\hat{\Pi}_{P4,\uparrow}^{\alpha}(\Omega,eV) \hat{\Pi}_{P1,\downarrow}^{\alpha}(-\Omega,eV) \Big] \nonumber\\
&&=4\int_{0}^{eV} \frac{d\Omega}{2\pi} \Big[\hat{\Pi}_{P2,\uparrow}^{\alpha}(\Omega,eV) \hat{\Pi}_{P3,\downarrow}^{\alpha}(-\Omega,eV)
                                -\hat{\Pi}_{P1,\uparrow}^{\alpha}(\Omega,eV) \hat{\Pi}_{P4,\downarrow}^{\alpha}(-\Omega,eV) \Big].
\end{eqnarray}

We want to expand the following integral in order of $eV$
\begin{eqnarray}
&& \quad\,\int_{0}^{eV} \frac{d\Omega}{2\pi} \hat{\Pi}_{Pi,\uparrow}^{\alpha}(\Omega,eV) \hat{\Pi}_{Pj,\downarrow}^{\alpha}(-\Omega,eV)\nonumber\\
 &&=\frac{1}{2 \pi} (eV) \hat{\Pi}_{Pi,\uparrow}^{\alpha}(0^+,eV) \hat{\Pi}_{Pj,\downarrow}^{\alpha}(0^-,eV) \nonumber\\
 &&\quad+ \frac{1}{2 \pi} (eV)^2 \frac{1}{2!} \Big( \frac{\partial \hat{\Pi}_{Pi,\uparrow}^{\alpha}(0^+,eV)}{\partial\Omega} \hat{\Pi}_{Pj,\downarrow}^{\alpha}(0^-,eV)
    -\hat{\Pi}_{Pi,\uparrow}^{\alpha}(0^+,eV)  \frac{\partial \hat{\Pi}_{Pj,\downarrow}^{\alpha}(0^-,eV)}{\partial\Omega} \Big)\nonumber\\
&& \quad+ \frac{1}{2 \pi} (eV)^3 \frac{1}{3!} \frac{\partial \hat{\Pi}_{Pi,\uparrow}^{\alpha}(\Omega,eV) \hat{\Pi}_{Pj,\downarrow}^{\alpha}(-\Omega,eV)}{\partial \Omega}\Bigg|_{\Omega\rightarrow 0^+} + \cdots
\end{eqnarray}
In the next step, we will expand the functions $\hat{\Pi}_{Pi,\sigma}^{\alpha}(0^{\pm},eV)$
$\partial_{\Omega}\hat{\Pi}_{Pi,\sigma}^{\alpha}(0^{\pm},eV)$ and $\partial^2_{\Omega}\hat{\Pi}_{Pi,\sigma}^{\alpha}(0^{\pm},eV)$
in the order of $eV$. The way to expand $\hat{\Pi}_{Pi,\sigma}^{\alpha}(0^{\pm},eV)$ can be found in appendix \ref{app:Derivation_XiA}.
For the $\partial_{\Omega}\hat{\Pi}_{Pi,\sigma}^{\alpha}(0^{\pm},eV)$, for example, we notice that
\begin{eqnarray}
 \frac{\partial  \hat{\Pi}_{P2}^{\alpha}(\Omega,eV)}{\partial\Omega}&=&\frac{i}{2\pi}\frac{\partial}{\partial_\Omega}
     \int_{-\infty}^{\infty} F_{2}^{\alpha}(\omega_1,\Omega|n_{R}(\omega_1+\Omega),n_{R}(\omega_1),n_{L}(\omega_1+\Omega),n_{L}(\omega_1)) d\omega_1\nonumber\\
     &=& \frac{i}{2\pi}\int_{-\infty}^{\infty}\Big\{ \partial_{\Omega}F_{2}(\omega_1,\Omega|0,0,0,0)
          +\big[\partial_{\Omega}F_{2}(\omega_1,\Omega|1,1,1,1)- \partial_{\Omega}F_{2}(\omega_1,\Omega|0,1,1,1)\big]n_{R}(\omega_1+\Omega)\nonumber\\
     &&\quad\quad\quad\quad\quad\quad\quad\quad\quad\quad\quad\quad +\big[\partial_{\Omega}F_{2}(\omega_1,\Omega|0,1,1,1)- \partial_{\Omega}F_{2}(\omega_1,\Omega|0,0,1,1)\big]n_{R}(\omega_1)\nonumber\\
     &&\quad\quad\quad\quad\quad\quad\quad\quad\quad\quad\quad\quad +\big[\partial_{\Omega}F_{2}(\omega_1,\Omega|0,0,1,1)- \partial_{\Omega}F_{2}(\omega_1,\Omega|0,0,0,1)\big]n_{L}(\omega_1+\Omega)\nonumber\\
     &&\quad\quad\quad\quad\quad\quad\quad\quad\quad\quad\quad\quad +\big[\partial_{\Omega}F_{2}(\omega_1,\Omega|0,0,0,1)- \partial_{\Omega}F_{2}(\omega_1,\Omega|0,0,0,0)\big]n_{L}(\omega_1) \Big\} d\omega_1 \nonumber\\
     &&\quad\quad -\frac{i}{2\pi}\Big[ F_{2}(-\Omega-\frac{eV}{2},\Omega|1,1,1,1)- F_{2}(-\Omega-\frac{eV}{2},\Omega|0,1,1,1)  \Big] \nonumber\\
     &&\quad\quad -\frac{i}{2\pi}\Big[ F_{2}(-\Omega+\frac{eV}{2},\Omega|0,0,1,1)- F_{2}(-\Omega+\frac{eV}{2},\Omega|0,0,01,1)  \Big].
\end{eqnarray}
Similarly, for $\partial_{\Omega}\hat{\Pi}_{Pj}^{\alpha}(\Omega,eV)$ with $j=1,3,4$.
Then, to expand the whole function, we just need to expand the Fermi distribution function and the function $F_{2}$ in order of $eV$, which is very
straightforward. For the second derivative $\partial^2_{\Omega}\hat{\Pi}_{Pi,\sigma}^{\alpha}(0^{\pm},eV)$, we can apply the
same strategy. Combing all the terms together, we recover Eq. \eqref{eq:XiB}.

\end{widetext}

\bibliography{ShotNoiseKondoMF,topological_wires11,FCSref}

\end{document}